\documentclass[twocolumn,amsmath,showpacs,amsfonts,aps,prc,floatfix,%
               groupedaddress]{revtex4}

\usepackage{epsfig}
\usepackage{bm}
\usepackage{natbib}

\newcommand{\2}{\frac{1}{2}}
\newcommand{\half}{{\textstyle\2}}

\newcommand\beq{\begin{equation}}
\newcommand\eeq{\end{equation}}
\newcommand\bea{\begin{eqnarray}}
\newcommand\eea{\end{eqnarray}}

\newcommand{\bq}{\bm{q}}
\newcommand{\bK}{\bm{K}}
\newcommand{\bp}{\bm{p}}

\newcommand{\Eq}{{\,=\,}}

\newcommand{\Ro}{R^2_o}
\newcommand{\Rs}{R^2_s}
\newcommand{\Rl}{R^2_l}
\newcommand{\Ri}{R^2_i}

\newcommand{\Cq}{C(\bq)}
\newcommand{\Cqk}{C\bigl(\bq^{(k)}\bigr)}
\newcommand{\qDotR}{\left(q_o^2R_o^2{+}q_s^2R_s^2{+}q_l^2R_l^2\right)}

\newcommand{\lCm}{\ln{\bigl(\Cqk-1\bigr)}}
\newcommand{\lnl}{\ln{\lambda}}
\newcommand{\csq}{\chi^2}
\newcommand{\sbin}{\sum_{k=1}^{N}}
\newcommand{\sprimeSq}{\left( \sigma_k^\prime \right)^2}

\newcommand{\qmax}{q_\mathrm{max}}

\begin{document}


\title{Fitted HBT radii versus space-time variances in flow-dominated models}

\author{Evan Frodermann}
\author{Ulrich Heinz}
\author{Michael Annan Lisa}
\affiliation{Physics Department, The Ohio State University, Columbus, OH 43210}

\begin{abstract} 
The inability of otherwise successful dynamical models to reproduce
the ``HBT radii'' extracted from two-particle correlations measured
at the Relativistic Heavy Ion Collider (RHIC) is known as the ``RHIC HBT
Puzzle''. Most comparisons between models and experiment exploit the fact 
that for Gaussian sources the HBT radii agree with certain combinations 
of the space-time widths of the source which can be directly computed 
from the emission function, without having to evaluate, at significant 
expense, the two-particle correlation function. We here study the 
validity of this approach for realistic emission function models some 
of which exhibit significant deviations from simple Gaussian behaviour. 
By Fourier transforming the emission function we compute the 2-particle 
correlation function and fit it with a Gaussian to partially mimic the
procedure used for measured correlation functions. We describe a
novel algorithm to perform this Gaussian fit analytically. We find 
that for realistic hydrodynamic models the HBT radii extracted from this 
procedure agree better with the data than the values previously 
extracted from the space-time widths of the emission function. Although 
serious discrepancies between the calculated and measured HBT radii 
remain, we show that a more ``apples-to-apples'' comparison of models 
with data can play an important role in any eventually successful
theoretical description of RHIC HBT data.
\end{abstract}


\date{\today}
\maketitle

\section{Introduction}
\label{sec1}

Two-particle intensity interferometry is widely used to characterize the 
space-time aspects of the freeze-out configuration in relativistic heavy 
ion collisions~\cite{Lisa:2005dd}. It is common to condense this 
information in terms of characteristic length scales of the ``homogeneity 
regions''~\cite{Akkelin:1995gh} from which particles of a given momentum 
originate.

In this paper we discuss the degree to which homogeneity lengths extracted
in quite different ways may be validly compared. Throughout our study, 
we restrict ourselves to interference effects between identical,
non-interacting bosons, resulting from Bose-Einstein statistics.
Since final state interactions (e.g. Coulomb effects) affect most 
interferometry studies, our study may be regarded (1) as a 
proof-of-principle example that care must be taken to perform 
``apples-to-apples'' comparisons, and (2) as an estimate of the 
magnitude of the differences for two popular theoretical models. 

The homogeneity length scales are extracted in experiments by assuming 
that the homogeneity region can be approximated by a Gaussian-profile 
ellipsoid in configuration space, resulting in a Gaussian two-particle 
momentum correlation function, and performing a semi-analytic Gaussian 
fit to the relative momentum dependence of the measured correlation 
function (see e.g. \cite{Lisa:2005dd} for details). Following common 
practice, we will refer in the following to the size parameters obtained 
from Gaussian fits to the correlation function as ``HBT radii''.

Fitting experimental data to functional forms other than Gaussian is 
common in studies of elementary particle collisions, for which Gaussian 
fits clearly fail. In heavy ion collisions, the Gaussian ansatz works 
relatively well, but, especially with the high quality and high-statistics
data sets now available at RHIC, finer, non-Gaussian structures may be 
physically interesting. Instead of inventing ad-hoc functional forms 
with which to fit the correlation functions, or functionally expanding 
about a Gaussian fitting form~\cite{Wiedemann:1996ej,Adams:2004yc}, 
imaging \cite{Brown:1997ku,Brown:2004bh,Brown:2005ze} the homogeneity 
region is perhaps the most promising route to explore these structures.
In this paper we do not take up this issue. Instead, we note that most 
experimental studies in heavy-ion physics to date have used the 
Gaussian ansatz~\cite{Lisa:2005dd}, and we explore some ways in which 
HBT radii obtained in this way from data may be compared to model 
calculations.

If the homogeneity region is indeed Gaussian in profile, then the HBT 
radii agree exactly with appropriate combinations of the 
root-mean-squared (RMS) variances of its spatial distribution
\cite{Wiedemann:1999qn}. Given a theoretical model for the freeze-out 
configuration, calculating these space-time variances is much easier 
than computing and fitting the correlation function. Many comparisons
between models and data therefore use this short-cut, comparing the 
space-time variances directly to the experimental HBT radii. However,
since the homogeneity region is seldom perfectly Gaussian, such direct 
comparisons are questionable. 

This raises the question to what extent some of the persistently observed 
discrepancies between model predictions and measurements of the HBT radii 
\cite{Lisa:2005dd} (the so-called ``RHIC HBT Puzzle'') might be due to 
such an ``apples-with-oranges'' comparison. Indeed, HBT radii calculated with 
Boltzmann/cascade models which are based on Gaussian fits to the simulated
correlation functions agree somewhat better with measurements than do radii
based on an extraction of space-time variances from hydrodynamic 
calculations~\cite{Lisa:2005dd}. Whether this is due to a more realistic 
modeling of the collision in the Boltzmann/cascade approach or the
shortcomings of the comparison of variances with HBT radii in the 
hydrodynamic case is unclear. 
Similarly, differences between hydrodynamic calculations of space-time variances~\cite{Kolb:2003dz,Heinz:2002un}
and Gaussian HBT radii fitted to three-dimensional~\cite{Hirano:2001yi,Morita:2002av,Morita:2003mj}
and one-dimensional~\cite{fn1,Hirano:2002ds} correlations have been observed.  However, since these
calculations were performed using different initial conditions and other parameters, it is unclear
whether this, or the different extraction methods, were responsible for the observed differences.
Here, we focus on the different extraction techniques using the same hydrodynamic model and parameters.

One cascade model (MPC \cite{Molnar:2002bz})
which reports RMS variances shows discrepancies with data similar to the 
hydrodynamic models. Studies~\cite{Hardtke:1999vf,Soff:2001hc,Lin:2002gc}
performed within the Boltzmann/cascade framework show that space-time variances of
the freeze-out configuration and Gaussian fits to the correlator can yield 
quite different radius parameters, mostly due to long tails in the spatial
freeze-out distribution from resonance decays which strongly affect the 
space-time variances but are not reflected by Gaussian fits to the correlation
function, according to hydrodynamic calculations~\cite{Wiedemann:1996ig}. 
(See, however, the recent study by Kisiel {\it et al}~\cite{Kisiel:2006is}, which addresses
this issue in detail in the context of a blast-wave parameterization.)
Hydrodynamic calculations of the 
space-time variances therefore usually do not include resonance decay contributions
in the emission function \cite{Kolb:2003dz}. Still, the comparison in
\cite{Kolb:2003dz} involves two differently determined quantities, and
in the present paper we eliminate this shortcoming.  
  
To do so requires two additional steps beyond the calculation of the
model emission function: (i) The correlation function must be computed 
via Fourier transformation (for noninteracting identical particles) 
or by folding with a relative wave function that includes final state 
interaction effects (for with long-range final state interactions). 
This is straightforward albeit numerically expensive since it involves 
multiple space-time integrals. (ii) A Gaussian fit to the 
three-dimensional correlation function must be performed, including a 
correlation strength parameter $\lambda$ as in the experiment. 

We here concentrate on non-interacting pairs of identical particles 
as the practically most important case and also in order to simplify 
as much as possible the computation of the correlator. For the second 
step we develop an analytical Gaussian fit algorithm which reduces 
the multi-dimensional fit problem to a simple set of linear equations 
for diagonalizing a four-dimensional matrix. This should help 
theoretical modelers to overcome the barrier of unfamiliarity when 
faced with a multi-parameter fitting problem. 

We apply our procedure to emission functions from hydrodynamic 
calculations \cite{Kolb:2003dz} and from the blast-wave parameterization 
\cite{Retiere:2003kf}. Both generate non-Gaussian freeze-out distributions,
due in large measure to finite-size effects coupled with strong collective 
flow which is known to be important at RHIC. On the way, we also discuss 
and analyze Gaussian fits to 1-dimensional projections of the 3-dimensional
correlator. This allows for comparison with earlier work along these
lines \cite{Wiedemann:1996ig,fn1} and first introduces our new analytic 
Gaussian fit algorithm in an easy and transparent simpler setting.  

\section{Variances versus HBT radii}
\label{sec2}

Experimentally, the correlation function between two iden\-ti\-cal particles, 
as a function of their relative mo\-men\-tum $\bq{\,\equiv\,}\bp_a{-}\bp_b$
and their average (pair) momentum $\bK{\,\equiv\,}(\bp_a{+}\bp_b)/2$, is 
given by
\begin{equation}
\label{eq:ExperimentalDefinition}
C(\bq,\bK) = \frac{A(\bq,\bK)}{B(\bq,\bK)},
\end{equation}
where $A(\bq,\bK)$ is the signal distribution and $B(\bq,\bK)$ is the 
reference or background distribution which is ideally similar to $A$ 
in all respects except for the presence of femtoscopic correlations 
(see e.g. \cite{Lisa:2005dd} for details). $C(\bq,\bK)$ is the
modification to the conditional probability for measuring particle 
$b$ with momentum $\bp_b\Eq\bK{-}\half\bq$ if particle $a$ has been 
measured with momentum $\bp_a\Eq\bK{+}\half\bq$, due to two-particle 
effects sensitive to space-time separation, to. The explicit 
$\bK$-dependence reflects the fact that the separation distribution 
may depend on the average momentum of the pair \cite{Akkelin:1995gh} 
and in general does so for exploding sources \cite{Pratt:1984su}.

Theoretically, the correlation function can be calculated from the 
emission function $S(\bp,x)$ describing the probability to emit a 
particle from spacetime point $x$ with momentum $\bp$, by convoluting 
it with the two-particle relative wave function \cite{Lisa:2005dd}. 
For pairs of non-interacting identical particles one has simply
\cite{Lisa:2005dd,Wiedemann:1996ej}
\beq
\label{eq:TheoryDefinition}
C(\bq,\bK) \approx 1 + \left| 
\frac{\int d^4x\, S(\bK,x)\,e^{iq{\cdot}x}}
     {\int d^4x\, S(\bK,x)}\right|^2.
\eeq
Here $q\cdot x\Eq{q^0}t-\bq\cdot\bm{x}$, with 
$q^0\Eq{E}_a{-}E_b\Eq\bm{\beta}\cdot\bq$ where 
$\bm{\beta}\Eq\bK/K^0\Eq2\bK/(E_a{+}E_b)$ is the average velocity
of the pair. The $\approx$ sign in Eq.~(\ref{eq:TheoryDefinition})
indicates the ``smoothness approximation'' which replaces both 
$\bp_a$ and $\bp_b$ by $\bK$ inside the emission functions in the 
denominator \cite{Wiedemann:1996ej}. Equation (\ref{eq:TheoryDefinition}) 
can be decomposed as
\beq
\label{eq:TheoryIdOnly}
C(\bq,\bK) = 1 + \langle\cos(q\cdot x)\rangle^2 
               + \langle\sin(q\cdot x)\rangle^2
\eeq
where $\langle\dots\rangle$ indicates the ($\bK$-dependent) space-time
average with the emission function:
\beq
\langle f \rangle \equiv \frac{\int d^4x\, f(x)\,S(\bK,x)}
                              {\int d^4x\, S(\bK,x)}\,.
\eeq

If $S(\bK,x)$ is a four-dimensional {\em Gaussian} distribution of freeze-out 
points, the correlation function will likewise be Gaussian in the relative 
momentum $\bq$. It takes a particularly simple form for midrapidity pairs
(with vanishing longitudinal pair momentum, $K_L\Eq0$) from central 
collisions between equal-mass spherical nuclei 
\cite{Lisa:2005dd,Wiedemann:1999qn}:
\beq
\label{eq:3dGauss}
\Cq = 1 + \lambda \, e^{-\qDotR }.
\eeq
Here $q_o,\,q_s,\,q_l$ are the relative momentum components in the 
Bertsch-Pratt (``out-side-long'') coordinate system 
\cite{Lisa:2005dd,Wiedemann:1999qn}. The pair momentum dependence of 
the correlation function $C(\bq,\bK)$ leads to $\bK$-dependencies 
of the ``HBT radii'' $R_o$, $R_s$, and $R_l$ (which characterize the 
relative momentum widths of the correlation function) and of the 
``correlation strength'' $\lambda$. For fully chaotic theoretical 
Gaussian sources $\lambda{\,\equiv\,}1$, but for experimental correlation 
functions usually $\lambda{\,<\,}1$. Even though we here perform a 
theoretical model analysis, we keep $\lambda$ as a parameter because 
Gaussian fits to non-Gaussian correlation functions generally also 
yield $\lambda{\,\ne\,}1$, and experimentally such non-Gaussian 
effects on the extracted $\lambda$ cannot be separated from other 
origins of reduced correlation strength (such as contamination from 
misidentified particles and contributions from resonance decays
\cite{Lisa:2005dd}). The HBT radii defined by Eq.~(\ref{eq:3dGauss}) 
convey all available geometric information about the source $S(\bK,x)$. 

For Gaussian sources the radius parameters $R_o$, $R_s$, and $R_l$ can be 
calculated directly from the source distribution $S$ as RMS variances. For 
midrapidity pairs with $K_L\Eq0$ one finds \cite{Wiedemann:1999qn}
\bea
\label{eq:VarianceRadii}
&&R^2_o = \langle \tilde{x}_o^2 \rangle 
      - 2\beta\langle\tilde{x}_o\tilde{t} \rangle 
      + \beta^2\langle \tilde{t}^2 \rangle, \nonumber \\
&&R^2_s = \langle \tilde{x}_s^2 \rangle, \qquad 
  R^2_l = \langle \tilde{x}_l^2 \rangle,
\eea
where $\beta\Eq{K_T}/K^0$ is the magnitude of the (transverse) pair 
velocity (which points in the $x_o$ direction), and
\beq
\tilde{x}^\mu \equiv x^\mu - \langle x^\mu \rangle
\eeq
denotes the distance from the ($\bK$-dependent) center of the homogeneity 
region for particles with momentum $\bK$.

Experimentalists commonly extract HBT radii by fitting their experimental 
correlation functions (\ref{eq:ExperimentalDefinition}) with the functional 
form (\ref{eq:3dGauss}). In contrast, most (but not all) theoretical 
model predictions for HBT radii are based on a calculation of the 
space-time variances of the emission function and assuming the validity
of Eqs.~(\ref{eq:VarianceRadii}) which holds for Gaussian sources.
Of course, there is no {\em a priori} reason to expect a source with 
a {\it perfectly} Gaussian profile. Even the simplest flow-dominated 
freeze-out parameterizations produce clear non-Gaussian tails and 
edges \cite{Retiere:2003kf}. On the experimental side, high-statistics 
measurements show non-Gaussian behaviour, which is, however rarely 
treated quantitatively \cite{Adams:2004yc}. In the presence of such
non-Gaussian features, the issues are (1) whether the two approaches 
yield significantly different results, and (2) whether either method 
characterizes the physically interesting length scales of the source
sufficiently well. Here, we address the first issue in the context of 
blast-wave and hydrodynamic models.

Our calculations do not include experimental ``noise'', particle 
mis-identification, or contributions from the decay of long-lived
resonances which can reduce the fit parameter $\lambda$ in 
Eq.~(\ref{eq:3dGauss}) from its theoretical value of unity 
\cite{Lisa:2005dd,Wiedemann:1996ig}. Instead, this parameter absorbs 
(and reflects) some of the effects of fitting a non-Gaussian function 
to a Gaussian form. This will, of course, also happen in experiment
whenever the correlation function deviates from a simple Gaussian.
This particular contribution to the fitted correlation strength $\lambda$ 
has so far received little attention. The model results presented here 
should help to assess the possible influence of non-Gaussian features 
in the data on the fitted values of $\lambda$.

\section{Direct calculation of HBT radii}
\label{sec3}

As explained in the Introduction, we here use model emission functions
to compute the correlation function according to 
Eqs.~(\ref{eq:TheoryDefinition},\ref{eq:TheoryIdOnly}) and then fit
the latter with a Gaussian, using a procedure very similar to the one
used in experiment. The main difference is that the theoretical correlation 
function can be calculated with arbitrary precision, so the notion of a 
statistical error does not enter. Still, we will see that the fitting
problem can be formulated in a quite analogous way.

In the following subsection we introduce the algorithm for Gaussian
fits through 1-dimensional cuts or projections of the 3-dimensional
correlation function. The full algorithm for 3-dimensional Gaussian
fits is presented in Sec.~\ref{sec3b}.

\subsection{One-dimensional Gaussian fits}
\label{sec3a}

In Section VI of their paper, Wiedemann and Heinz \cite{Wiedemann:1996ig} 
calculated correlators for various model emission functions and extracted 
parameters from fits to {\it one-dimensional slices} of the three-dimensional 
correlation function. Although those authors called them ``HBT radii'', 
we will call them ``1D radii'' to distinguish them from radii extracted
from full three-dimensional fits of the type performed by experimentalists.

In a given direction $i$ ($i = o,s,l$) they calculate the correlator 
along one of the axes $i$: $C(q_i; q_{j{\neq}i}{=}0)$. They then find 
the 1D radius $R^2_{\mathrm{1D},i}$ and the ``directional lambda 
parameter'' $\lambda_i$ which best approximates the correlator 
according to
\beq
\label{eq:WH1}
  C(q_i; q_{j{\neq}i}{=}0) \approx 1 + \lambda_i\, 
  e^{-q_i^2  R^2_{\mathrm{1D},i}} \, .
\eeq
In particular, they calculated the correlator for a set of $N$ values 
$q_i^{(k)}$ (similar to experimentally binning the correlation function 
into $N$ $\bq$-bins) and minimized numerically the quantity
\beq
\label{eq:WH2}
 \sbin \left[ \ln\bigl(C(q_i^{(k)};q_{j{\neq}i}^{(k)}{=}0) - 1\bigr)
            - \ln\lambda_i + R^2_{\mathrm{1D},i} (q_i^{(k)})^2 \right]^2 .
\eeq
This is reminiscent of the quantity typically minimized by experimenters,
although in this case one also takes into account the experimental 
uncertainty of the measured correlator by weighting each term in the
sum (bin) with the inverse experimental error:
\bea
\label{eq:chi2OD}
 &&\chi^2_{\mathrm{1D},i} \equiv 
 \\
 &&\sbin \left[\frac{
               \ln\bigl[C\bigl(q_i^{(k)};q_{j{\neq}i}^{(k)}{=}0\bigr){-}1\bigr]
             - \ln\lambda_i + R^2_{\mathrm{1D},i} (q_i^{(k)})^2}
              {{\sigma'}_{\mathrm{1D},i}^{\,(k)}}\right]^2.
 \nonumber
\eea
Here, ${\sigma'}_{\mathrm{1D},i}^{\,(k)}$ represents the uncertainty in 
bin $k$ on the quantity to be fitted, namely 
$\ln\bigl[C\bigl(q_i^{(k)};q_{j{\neq}i}^{(k)}{=}0\bigr)-1\bigr]$. It is 
related to the uncertainty $\sigma_{\mathrm{1D},i}^{(k)}$ on the
measured correlator $C(q_i^{(k)};q_{j{\neq}i}^{(k)}{=}0)$ itself by
\bea
\label{eq:sps}
  {\sigma'}_{\mathrm{1D},i}^{\,(k)} &=&
  \frac{d\ln\bigl[C\bigl(q_i^{(k)};q_{j{\neq}i}^{(k)}{=}0\bigr)-1\bigr]}
       {dC\bigl(q_i^{(k)};q_{j{\neq}i}^{(k)}{=}0\bigr)}
  \cdot \sigma_{\mathrm{1D},i}^{(k)}
\nonumber\\
  &=&
  \frac{\sigma_{\mathrm{1D},i}^{(k)}}
       {C\bigl(q_i^{(k)};q_{j{\neq}i}^{(k)}{=}0\bigr)-1}\,.
\eea
Minimization of the quantity (\ref{eq:WH2}) as in \cite{Wiedemann:1996ig} is 
equivalent to setting all uncertainties ${\sigma'}_{\mathrm{1D},i}^{\,(k)}$
to the same constant value, independent of $k$. However, uncertainties on 
experimental correlation functions typically have approximately constant 
($k$-independent) uncertainties on the bin contents 
$C(q_i^{(k)};q_{j{\neq}i}^{(k)}{=}0)$ themselves \cite{fn2}. 
Although statistical uncertainties on calculated correlators may in 
principle be vanishingly small, the weighting factor 
$\bigl[C\bigl(q_i^{(k)};q_{j{\neq}i}^{(k)}{=}0\bigr){-}1\bigr]^2$ which 
appears in Eq.~(\ref{eq:chi2OD}) as a result of Eq.~(\ref{eq:sps}) will 
in general affect the resulting fit parameters. We choose to mimic 
the experimental situation by minimizing Eq.~(\ref{eq:chi2OD}), assuming 
constant (i.e. $k$-independent) and infinitesimally small errors on $C$,
$\sigma_{\mathrm{1D},i}^{(k)}\Eq\sigma_{\mathrm{1D},i}{\,\to\,}0$.

Minimizing $\chi^2_{\mathrm{1D},i}$ in Eq.~(\ref{eq:chi2OD}) with respect
to the fit parameters $\ln\lambda_i$ and $R^2_{\mathrm{1D},i}$
by setting
\beq
 \frac{\partial \chi^2_{\mathrm{1D},i}}{\partial \ln\lambda_i} = 0 \,, 
 \qquad  
 \frac{\partial \chi^2_{\mathrm{1D},i}}{\partial R^2_{\mathrm{1D},i}} = 0 \,,
\eeq
we find after minimal algebra 
\bea
\label{eq:1Dlammu}
  \ln\lambda_i &=& \frac{X_{2,i} Y_{2,i} - X_{0,i} Y_{4,i}}
                        {Y^2_{2,i} - Y_{0,i} Y_{4,i}}\,,
\\
\label{eq:1DRmu}
  R^2_{\mathrm{1D},i} &=& 
  \frac{X_{2,i} Y_{0,i} - X_{0,i} Y_{2,i}}
       {Y^2_{2,i} - Y_{0,i} Y_{4,i}} \,,
\eea 
where the quantities
\bea
\label{eq:X1D}
  X_{n,i} &=& \sbin 
  \frac{\bigl(q_i^{(k)}\bigr)^n}
       {\bigl({\sigma'}_{\mathrm{1D},i}^{\,(k)}\bigr)^2}\,
  \ln\bigl[C\bigl(q_i^{(k)};q_{j{\neq}i}^{(k)}{=}0\bigr) - 1\bigr],\quad         
\\
\label{eq:Y1D}
  Y_{n,i} &=& \sbin 
  \frac{\bigl(q_i^{(k)}\bigr)^n}
       {\bigl({\sigma'}_{\mathrm{1D},i}^{\,(k)}\bigr)^2}
\eea
are directly calculable from the calculated correlator. Note
that the constant error $\sigma_{\mathrm{1D},i}$ of the correlator 
drops out from the ratios in Eqs.~(\ref{eq:1Dlammu},\ref{eq:1DRmu}),
so the limit $\sigma_{\mathrm{1D},i}{\,\to\,}0$ mentioned above is 
well-defined.

Minimization of $\chi^2_{\mathrm{1D},i}$ differs significantly from the 
experimentalists' three-dimensional fits. In particular, it assumes 
complete factorization of the correlation function in the $o,s,l$ 
directions. For at least two reasons, this need not be so in reality:

(i) In a full three-dimensional fit, the three directions are coupled
by requiring a single $\lambda$ parameter, independent of direction $i$. 
After all, according to Eq.~(\ref{eq:WH1}) $\lim_{|\bq|\to0}
C(\bq)\Eq\lim_{q_i\to0}C(q_i;q_{j{\ne}i{=}0})\Eq1{\,+\,}\lambda_i$ 
should be independent of direction $i$. Thus, allowing ``directional 
lambda parameters'' may cause the 1D fits to differ significantly from 
3D fits.

(ii) Perhaps more importantly, fitting separately along the $q_i$ axes 
accounts for only a set of zero measure of the full three-dimensional 
correlation function. In particular, the correlation function may contain
in the exponent terms such as $q_o^2q_s^2$ or $q_o^4q_l^2$. 
(For symmetry reasons \cite{Heinz:2002au} odd powers of $q_i$ vanish at 
midrapidity for central collisions between equal nuclei.)
Such higher order terms will affect the 3D fits of the experimentalist, 
but have no effect on equation (\ref{eq:chi2OD}).

We therefore now turn to full three-dimensional Gaussian fits. We will
see that the above analytic expressions are easily generalized for this
case.

%
\begin{figure}[t]
\centerline{\includegraphics[bb=20 20 495 495,width=0.5\textwidth,clip=]%
                            {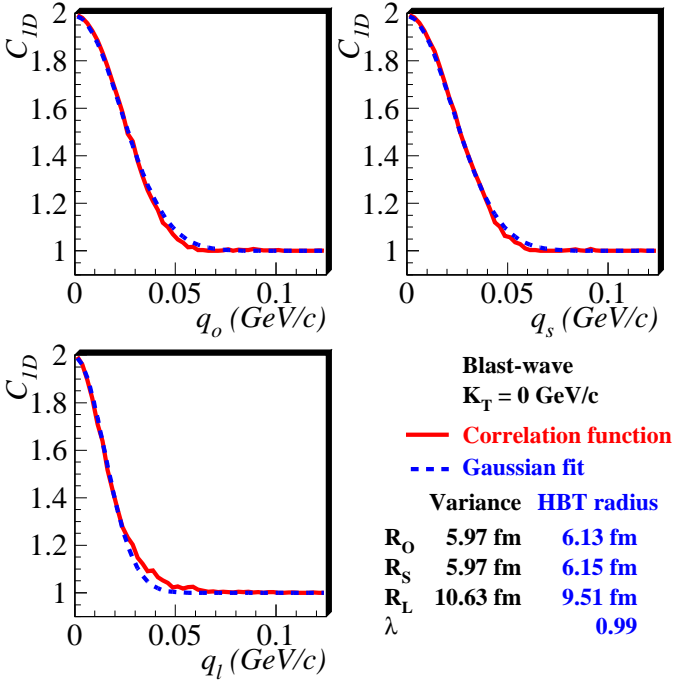}}
\caption{
\label{fig:1dProj-pT0.0}
(Color online) One-dimensional slices of the three-dimensional correlation 
function along the ``out'', ``side'', and ``long'' directions, for pion 
pairs with $\bK\Eq0$, calculated from the blast-wave parameterization.
For a given slice, the unplotted $q$-components equal 1.25~MeV/$c$ 
(i.e. the center of the first bin). The solid (red) curve is the 
calculated correlation function from Eq.~(\ref{eq:TheoryIdOnly}), the
dashed (blue) curve shows the same slice of the best 3D Gaussian fit 
(\ref{eq:3dGauss}), with ``HBT parameters'' calculated from the analytic 
expressions given in Sec.~\ref{sec3b}.
}
\end{figure}
%
%
\begin{figure}[t]
\centerline{\includegraphics[bb=20 20 495 495,width=0.5\textwidth,clip=]%
                            {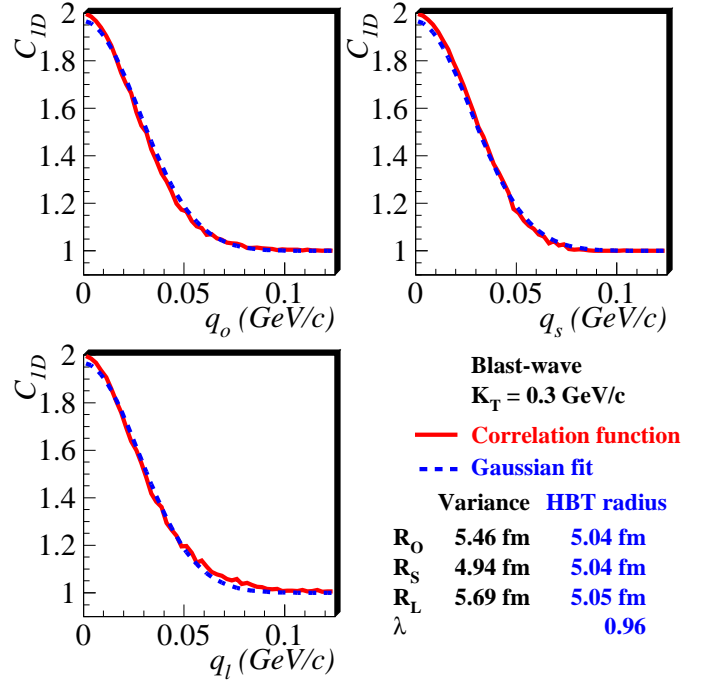}}
\caption{
\label{fig:1dProj-pT0.3}
(Color online) 
Solid (red) curves show
one-dimensional slices of the three-dimensional correlation function calculated
with Eq.~(\ref{eq:TheoryIdOnly}) from the blast-wave parameterization, for midrapidity
pions with $K_T=0.3$~GeV/$c$.  Dashed (red) curves show slices of the three-dimensional
Gaussian form of Equation~(\ref{eq:3dGauss}), with ``HBT parameters'' calculated from the analytic 
expressions given in Sec.~\ref{sec3b}.
}
\end{figure}
%

\subsection{Three-dimensional Gaussian fit algorithm}
\label{sec3b}

Proceeding as in the previous subsection, we start from the general 
three-dimensional Gaussian ansatz (\ref{eq:3dGauss}) which can be 
written as
\bea
\label{eq:logC}
 \ln\bigl(\Cq{-}1\bigr) = \ln\lambda - \qDotR .
\eea
If the correlation function $\Cqk$ in bin $k$ has error $\sigma_k$,
the error on $\ln(C{-}1)$ is given as in (\ref{eq:sps}) by
\beq
\label{eq:sigma_prime}
  \sigma'_k = \frac{\sigma_k}{\Cqk-1}\,.
\eeq
We minimize
%
%
%
\beq
\label{eq:chisq}
  \chi^2 = \sbin \left[
  \frac{\lCm-\lnl+{\displaystyle{\sum_{i=osl}}}\bigl(q_i^{(k)}\bigr)^2 \Ri}
       {\sigma'_k}\right]^2
\eeq
by setting
\beq
\label{eq:Partials}
\frac{\partial \csq}{\partial \lnl} = 0\,, \qquad
\frac{\partial \csq}{\partial \Ri} = 0 \quad (i=o,s,l)\,. 
\eeq
This leads to a set of 4 coupled linear equations,
\beq
\label{eq:LinearEqs}
   \sum_{\beta} T_{\alpha\beta} P_{\beta} = V_{\alpha} \,,
\eeq
where $\alpha$ and $\beta$ take the values $\mathrm{\o},o,s,l$. The 
vectors appearing here are
\bea
\label{eq:P}
  P &=& \left( \lnl , \Ro , \Rs , \Rl \right) ,
\\
  V_\mathrm{\o} &=& - \sbin\frac{\lCm}{\sprimeSq} \,,
\\
\label{eq:V}
  V_i & = & + \sbin\frac{\bigl(q_i^{(k)}\bigr)^2}
                            {\sprimeSq}\cdot\lCm ,\quad
\eea
while the symmetric $4\times 4$ matrix $T$ has components
\bea
\label{eq:T}
  T_{\mathrm{\o}\mathrm{\o}}  &=& - \sbin\frac{1}{\sprimeSq}\,,
\nonumber \\
  T_{\mathrm{\o}i}  &=& + \sbin\frac{\bigl(q_i^{(k)}\bigr)^2}{\sprimeSq}\,,
\\
  T_{ij} &=& - \sbin\frac{\bigl(q_i^{(k)}\bigr)^2\,\bigl(q_j^{(k)}\bigr)^2}
                          {\sprimeSq}\,.
\nonumber
\eea
In Equations~(\ref{eq:V}) and (\ref{eq:T}) $i,j\Eq{o,s,l}$ as usual. Note
the correspondences $V_\alpha \leftrightarrow X_{n,i}$ and
$T_{\alpha\beta}\leftrightarrow Y_{n,i}$ between the 3D and 1D
cases.

The set of linear equations (\ref{eq:LinearEqs}) is easily solved 
algebraically by diagonalizing the matrix $T_{\alpha\beta}$.

\section{Application to blast-wave model}
\label{sec4}

Many variants of ``hydrodynamically-inspired'' models of freeze-out have 
recently been used to calculate spatial RMS variances which then were 
compared to experimental HBT radii. A recent example is reported in 
reference~\cite{Retiere:2003kf}. The model itself is very simplistic 
and ignores, for example, resonance decay contributions which may be 
important \cite{Wiedemann:1996ig}. We ignore such issues with the model 
itself and simply use it here to discuss differences between RMS 
variances and Gaussian HBT radii.  

We use ``realistic'' model parameters which best describe the data 
\cite{Adams:2004yc}. Specifically, we take $R\Eq13.3$~fm for the source 
radius, $T\Eq97$~MeV for the temperature, $\rho_0\Eq1.03$ for the maximum 
transverse flow rapidity, $\tau\Eq9$~fm/c for the average freeze-out time, 
and $\Delta\tau\Eq2.83$~fm/c for the emission duration (see
\cite{Retiere:2003kf} for details).

\subsection{Correlation functions and analytic fits: results}
\label{sec4a}

Equation~(12) of~\cite{Retiere:2003kf} gives the functional form for 
the single-pion emission function in the blast-wave model. Using 
this for $S(\bK,x)$, we calculate the correlation function for pion
pairs with longitudinal pair momentum $K_L\Eq0$, using a Monte Carlo 
technique to numerically perform the integrals in Eq.~(\ref{eq:TheoryIdOnly}).

%
\begin{figure}[t]
\centerline{\includegraphics[bb=0 20 820 530,width=0.5\textwidth,clip=]%
                            {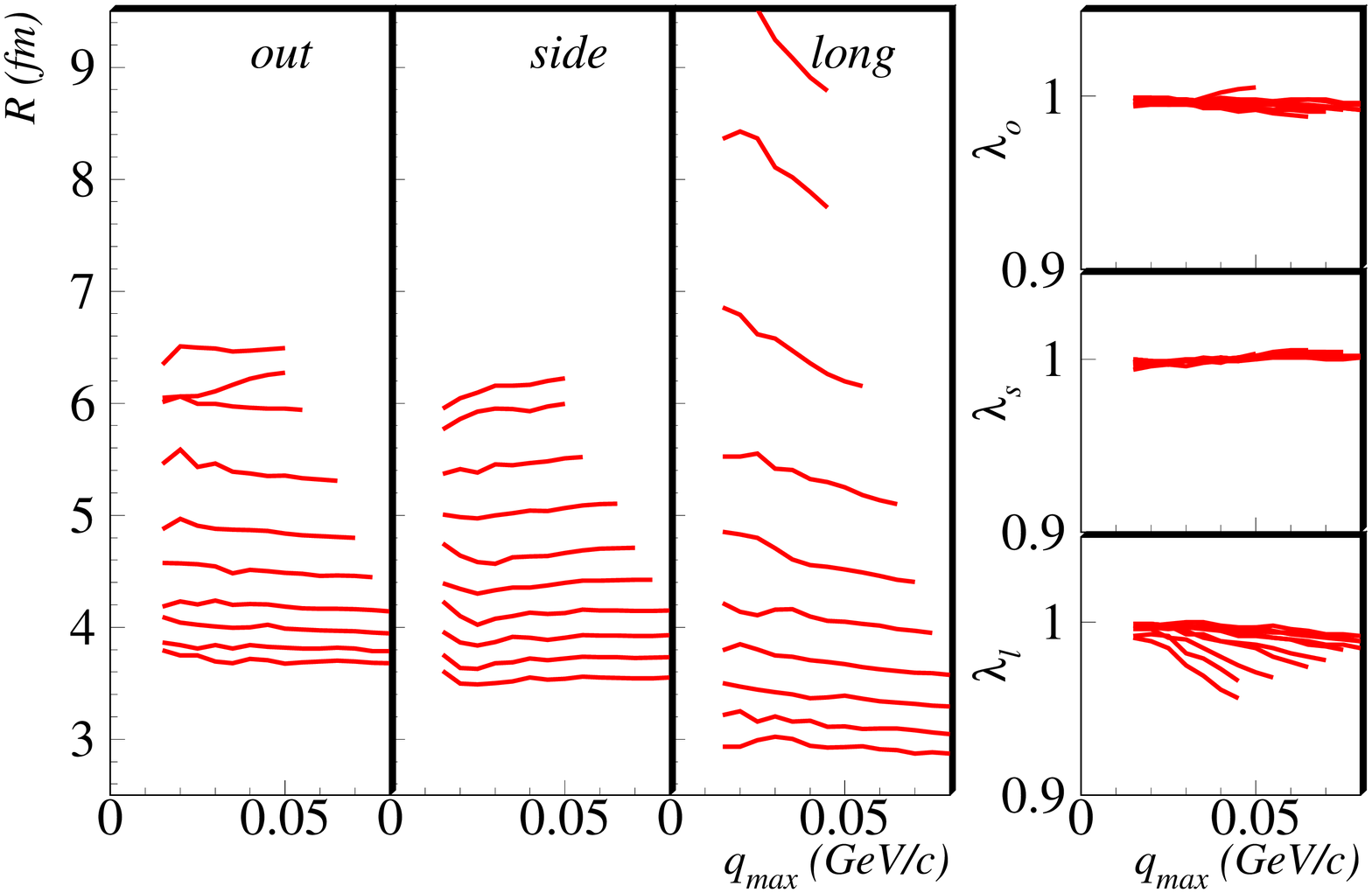}}
\caption{
\label{fig:fitrange1Dqmax}
(Color online)
From the blast-wave parameterization, one-dimensional HBT fit parameters 
$R_{\mathrm{1D},i}$ and $\lambda_{\mathrm{1D},i}$ are calculated with Eqs.~(\ref{eq:1Dlammu},\ref{eq:1DRmu}) 
and plotted as a function of the maximum allowed value of any $q$-component; see text   
for details. Each curve corresponds to one of ten values of $K_T$:
0.0, 0.1, 0.2, \dots, 0.9~GeV/$c$. Curves corresponding to high $K_T$ are 
at low (high) values of $R_{\mathrm{1D},i}$
($\lambda_{\mathrm{1D},i}$).
\vspace*{-3mm}
}
\end{figure}
%
%
\begin{figure}[t]
\centerline{\includegraphics[bb=0 20 820 530,width=0.5\textwidth,clip=]%
                            {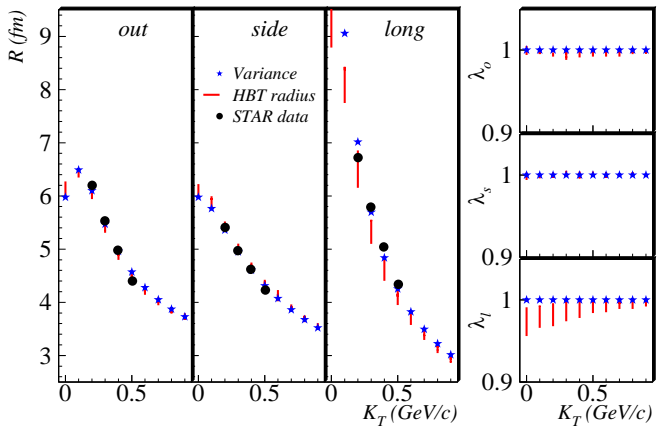}}
\caption{
\label{fig:fitrange1DpT}
(Color online) One-dimensional HBT fit parameters $R_{\mathrm{1D},i}$ and 
$\lambda_{\mathrm{1D},i}$ as a function of $K_T$, calculated from the
blast-wave parametrization with Eqs.~(\ref{eq:1Dlammu},\ref{eq:1DRmu}).
For a given $K_T$, the vertical red line represents the variation with 
fit range (see Figure~\ref{fig:fitrange1Dqmax}). Blue stars represent 
the corresponding radius parameters calculated from the RMS variances 
using Eq.~(\ref{eq:VarianceRadii}). Black circles show STAR data
\cite{Adams:2004yc}, with error bars removed for clarity.
\vspace*{-3mm}
}
\end{figure}
%

As with experimental data, the correlation function is evaluated in 
finite-sized three-dimensional bins in $(q_o,q_s,q_l)$ of width
2.5~MeV/$c$ in each direction. One-dimensional slices of the correlation 
function in the ``out'', ``side'', and ``long'' directions are
shown in Figures \ref{fig:1dProj-pT0.0} and \ref{fig:1dProj-pT0.3},
for midrapidity pion pairs with $K_T\Eq0$ and $K_T\Eq0.3$\,GeV/$c$, 
respectively.

%
\begin{figure}[t]
\centerline{
\includegraphics[bb=0 20 820 530,width=0.5\textwidth,height=5.58cm,clip=]%
                {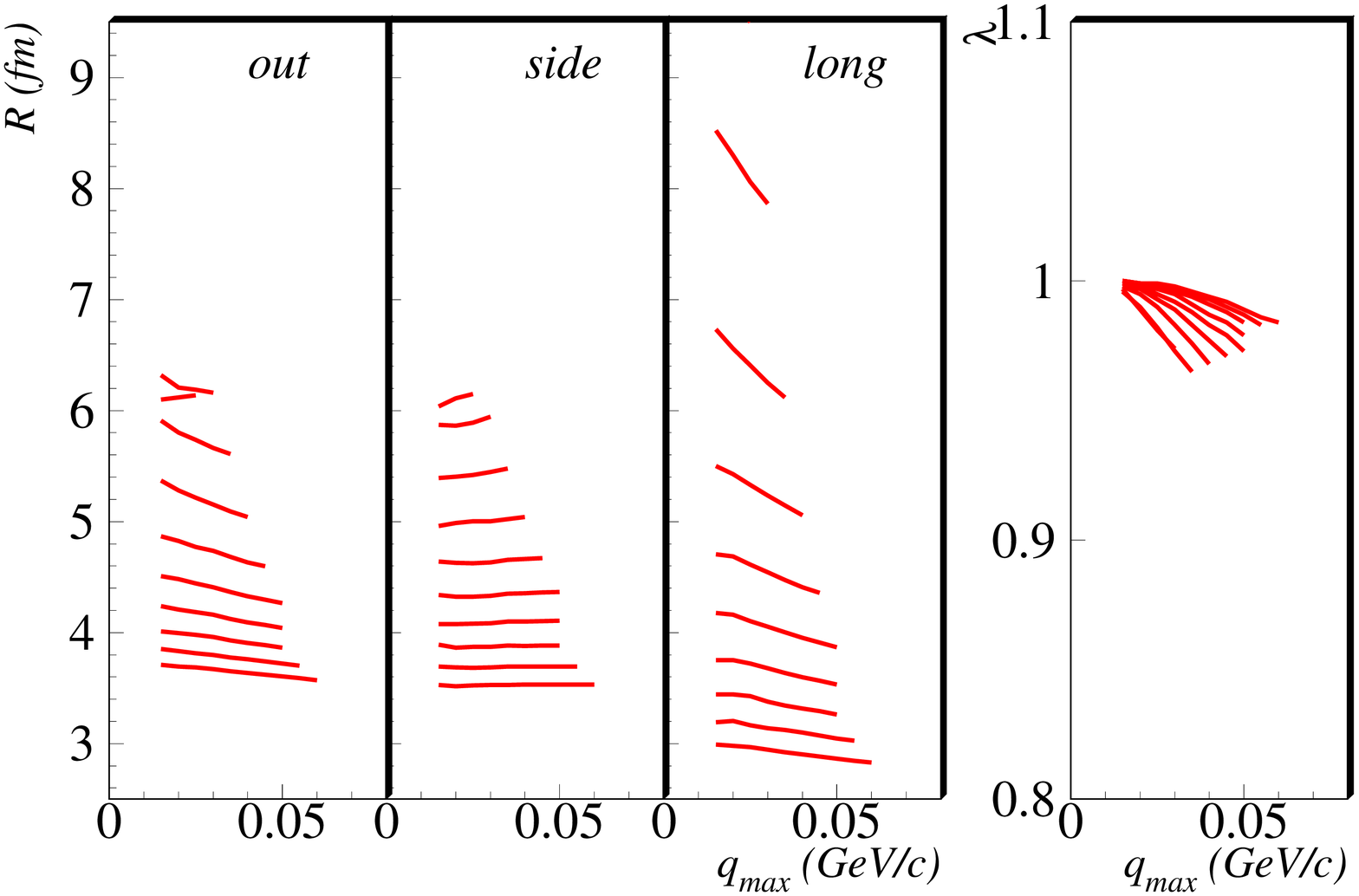}}
\caption{
\label{fig:fitrange3Dqmax}
(Color online)
From the blast-wave parameterization, three-dimensional HBT fit parameters 
$R_{i}$ and $\lambda$ are calculated with Eqs.~(\ref{eq:LinearEqs}) 
and plotted as a function of the maximum allowed value of any $q$-component; see text   
for details. Each curve corresponds to one of ten values of $K_T$:
0.0, 0.1, 0.2, \dots, 0.9~GeV/$c$. Curves corresponding to high $K_T$ are 
at low (high) values of $R_{i}$
($\lambda$).
The $R_l$ curve for $K_T\Eq0$ falls above
the plotting range.
}
\end{figure}
%
%
\begin{figure}[t]
\centerline{
\includegraphics[bb=0 20 820 530,width=0.5\textwidth,height=5.58cm,clip=]%
                {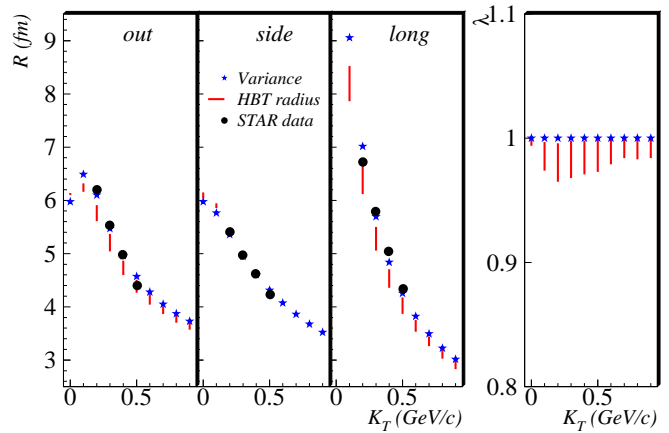}}
\caption{
\label{fig:fitrange3DpT}
(Color online) Three-dimensional HBT fit parameters $R_{\mathrm{1D},i}$ and 
$\lambda_{\mathrm{1D},i}$ as a function of $K_T$, calculated from the
blast-wave parametrization with Eqs.~(\ref{eq:LinearEqs}).
For a given $K_T$, the vertical red line represents the variation with 
fit range (see Figure~\ref{fig:fitrange3Dqmax}). Blue stars represent 
the corresponding radius parameters calculated from the RMS variances 
using Eq.~(\ref{eq:VarianceRadii}). Black circles show STAR data
\cite{Adams:2004yc}, with error bars removed for clarity.
\vspace*{-3mm}
}
\end{figure}
%

The slices of the correlation functions appear quite Gaussian, and they 
are tracked well by the three-dimensional Gaussian fit; the fitted
correlation strength $\lambda$ is very close to 1. The radius parameters 
calculated from the RMS variances (\ref{eq:VarianceRadii}) agree quite 
well with the HBT radii extracted from the three-dimensional Gaussian 
fit by solving Eqs.~(\ref{eq:LinearEqs}); both sets are given in the 
Figures. Upon closer inspection one notices, however, that the fitted 
outward and longitudinal radii, $R_o$ and especially $R_l$, tend to be 
systematically smaller than those extracted from the spatial RMS 
variances; the opposite is true for the sideward radii $R_s$ for 
which the RMS variances give slightly smaller values than the Gaussian 
fit. While these differences are small for the blast-wave model 
parameterization (at least with the ``realistic'' parameters studied 
here), they will be significantly larger (with the same basic tendencies 
as found here) for the hydrodynamic model source studied in 
Sec.~\ref{sec5}.

The Gaussian fit parameters given in Figs.~\ref{fig:1dProj-pT0.0} and 
\ref{fig:1dProj-pT0.3} correspond to using the largest possible 
$\bq$-range in the sums over $k$ in Eqs.~(\ref{eq:V},\ref{eq:T}), 
discarding only those data points for which $C$ is so close to 1
that the Monte Carlo integration sometimes yields negative values
for $C{-}1$. Due to small but noticeable deviations of the correlation 
function from a pure Gaussian, the Gaussian fit parameters depend on 
the number of data points used. We study this sensitivity to the fit range 
in the following subsection.

\subsection{Fit-range study}
\label{sec4b}

Since no measured correlation function is ever perfectly Gaussian,
experimentalists typically perform so-called ``fit range studies.'' 
Here, the measured correlation function is fitted with the Gaussian 
form (\ref{eq:3dGauss}), using data points in a restricted range of
$\bq$. With correlation functions in the one-dimensional quantity 
$Q_\mathrm{inv}$ it is common to study the variation of fit parameters 
as the first few (lowest-$Q_\mathrm{inv}$) data points are left out of 
the fit. This is because statistical fluctuations in these bins may be 
quite large, and due to the visible non-Gaussian nature of the measured 
correlation function there. Three-dimensional correlation functions do 
not suffer from these issues, and so usually the experimentalist includes 
all data points with $|q_i|{\,<\,}\qmax$ and studies variations of the
fit parameters as $\qmax$ is varied; any such variations are typically 
folded into systematic errors on the HBT radii.

Here, we follow the experimentalists' approach. Using the correlation 
function generated from the blast-wave model, we calculate HBT parameters 
from 1D and 3D Gaussian fits as discussed in Sections~\ref{sec3a} and
\ref{sec3b}, restricting the $k$-sums in Eqs.~(\ref{eq:X1D}), 
(\ref{eq:Y1D}), (\ref{eq:V}), and (\ref{eq:T}) to include only those 
data points where all three $q$-components have magnitudes less than 
$\qmax$ \cite{fn3}. Thus, we will not calculate unique HBT 
radii, but a finite range for each fit parameter.

For various values of $K_T$, Figure~\ref{fig:fitrange1Dqmax} shows
the evolution of the 1D radii with $\qmax$. Except for $R_{l}$ at low 
$K_T$, the parameter variation with fit range is quite mild, corresponding 
to a small ``non-Gaussian systematic error'' on the radii. In 
Figure~\ref{fig:fitrange1DpT} the range of this variation, indicated 
by vertical lines, is plotted as a function of $K_T$. Consistent with 
the theorem \cite{Wiedemann:1999qn} that the spatial RMS variances 
(\ref{eq:VarianceRadii}) of the source control the {\em curvature} of 
the correlator $C(\bq)$ at $\bq\Eq0$, the blue stars 
in Fig.~\ref{fig:fitrange1DpT} coincide with the $\qmax{\,\to\,}0$ 
limit of the fitted 1D radii. The largest fit-range variations,
indicating the biggest non-Gaussian effects in the correlator, are
seen at small pair momentum $K_T$. The fit-range sensitivity is most 
pronounced for $R_l$ (where at low $K_T$ it can exceed 0.5\,fm) 
but almost negligible for $R_o$ and $R_s$. In short, the 1D Gaussian 
fits to the two transverse projections of the correlation function give 
length scales consistent with the spatial RMS variances of the source 
distribution, but non-Gaussian features along the longitudinal projection 
cause the RMS variances to overestimate the longitudinal 1D HBT radius 
$R_l$ by up to 0.5\,fm at low $K_T$ if a reasonable fit range $\qmax$
is used to extract the latter. This discrepancy is significantly larger 
than the combined statistical and systematic error on the experimental
value for $R_l$ \cite{Adams:2004yc}.

Figures~\ref{fig:fitrange3Dqmax} and \ref{fig:fitrange3DpT} show 
the same study for the three-dimensional Gaussian fits. For reasons
explained in Sec.~\ref{sec3a}, the non-Gaussian effects in a unified 
3D Gaussian fit are expected to differ from those in 1D fits.
Indeed, in the unified 3D fit non-Gaussian influences also appear
in $R_o$, and both $R_o$ and $R_l$ now show fit-range variations 
which exceed the combined statistical and systematic errors of the 
data \cite{Adams:2004yc}. The largest fit-range sensitivity is still 
seen in the longitudinal direction. In Ref.~\cite{Retiere:2003kf} 
the blast-wave model parameters were determined by comparing RMS 
variances with the measured HBT radii (see Figs.~\ref{fig:fitrange1DpT} 
and \ref{fig:fitrange3DpT}), using the experimental errors on the 
latter to extract error estimates for the model parameters. The 
results presented here suggest that if the authors had instead 
compared the measured data with HBT radii extracted from a 3D Gaussian 
fit to the calculated correlation function, they would have found 
somewhat different model parameters whose mean values in some 
cases might even have fallen outside the likely parameter range 
quoted in Table~II of Ref.~\cite{Retiere:2003kf}. In particular,
such an ``apples-to-apples'' comparison may allow for somewhat 
larger fireball lifetimes $\tau$ and/or emission durations $\Delta\tau$ 
than quoted in Ref.~\cite{Retiere:2003kf}. While such an improved 
blast-wave model fit is numerically expensive and outside the scope 
of the present paper, it may be a worthwhile future project.

\section{HBT radii from hydrodynamics}
\label{sec5}

Non-viscous (``ideal'') hydrodynamical calculations have successfully 
reproduced differential momentum spectra (at least perpendicular to the 
beam direction) at RHIC, including their anisotropies in non-central
collisions and the dependence of these anisotropies on the masses of
the emitted hadrons \cite{Kolb:2003dz}. As in the blast-wave model 
calculations, very strong collective flow is a critical ingredient to 
reproduce the data. (Of course, in the blast-wave parameterization 
such flow is put in by hand while it arises naturally in the 
hydrodynamical model.)

Most (but not all~\cite{Hirano:2001yi,Hirano:2002ds,Morita:2002av,Morita:2003mj})
hydrodynamic predictions of HBT radius parameters have been based 
on calculations of the spatial RMS variances from the hydrodynamically 
generated emission function, using Eqs.~(\ref{eq:VarianceRadii}) 
\cite{Kolb:2003dz,Heinz:2002un}. 
In spite of the hydrodynamic 
model's impressive success in describing hadron spectra, these 
predictions of HBT radii were a failure: The calculated longitudinal 
radii $R_l$ were too large (although this problem was less severe in 
Hirano and Tsuda's work \cite{Hirano:2002ds}), while the predicted 
sideward radius $R_s$ was too small, and both $R_s$ and $R_o$ showed 
much less dependence on $K_T$ in theory than seen in the data. This, 
together with similar failures by other dynamical models (see 
\cite{Lisa:2005dd} for a review), has become known as the ``RHIC 
HBT Puzzle''. 

%
\begin{figure}[t]
\centerline{\includegraphics[bb=20 20 500 495,width=0.5\textwidth,clip=]%
                            {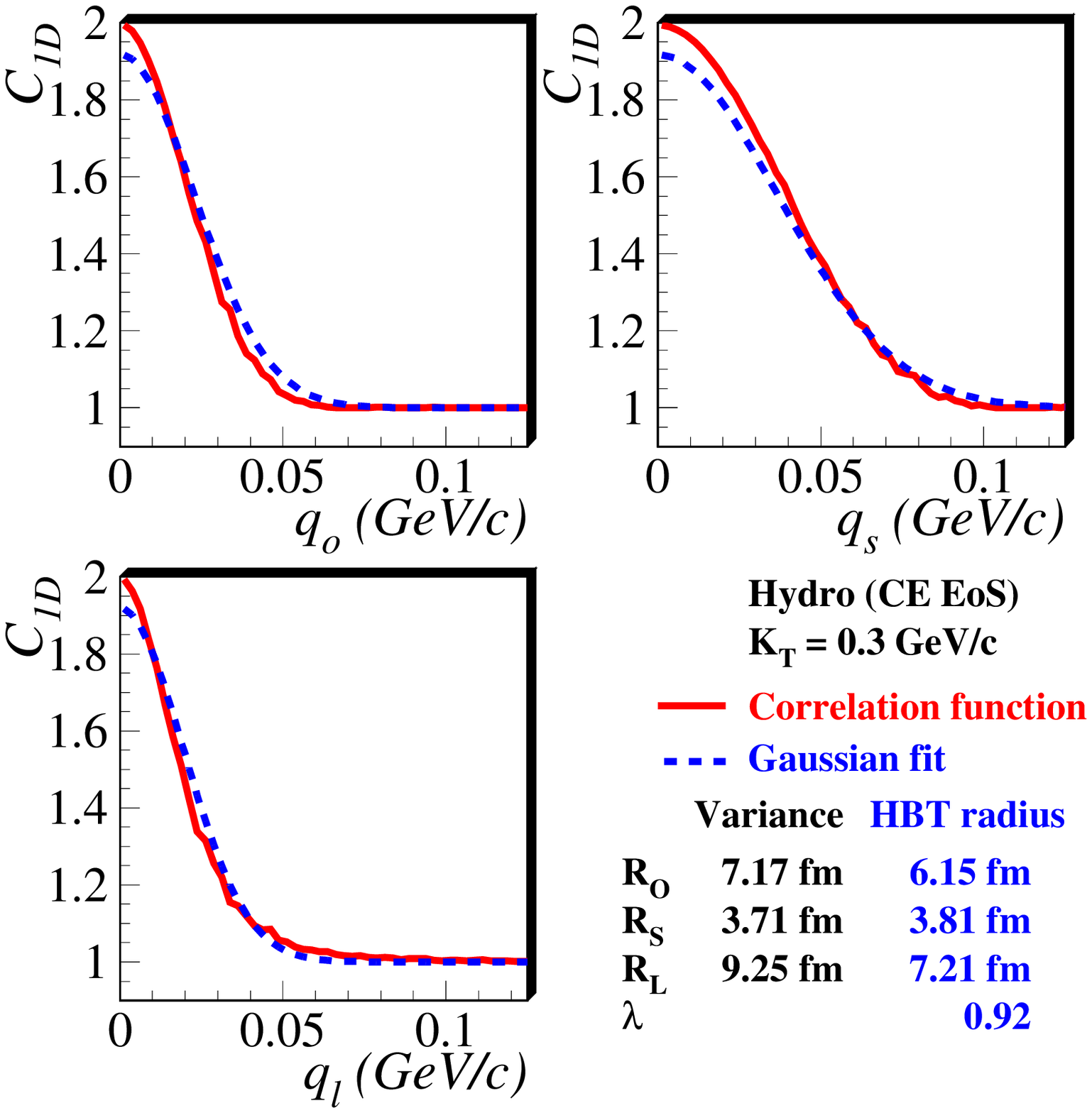}}
\caption{
\label{fig:CE1dProj-pT0.3}
(Color online) 
Solid (red) curves show
one-dimensional slices of the three-dimensional correlation function calculated
with Eq.~(\ref{eq:TheoryIdOnly}) from the hydrodynamic model with CE Equation of State, for midrapidity
pions with $K_T=0.3$~GeV/$c$.  Dashed (red) curves show slices of the three-dimensional
Gaussian form of Equation~(\ref{eq:3dGauss}), with ``HBT parameters'' calculated from the analytic 
expressions given in Sec.~\ref{sec3b}.
\vspace*{-1mm}
}
\end{figure}
%

Various possibilities to explain and correct this failure have been
suggested. They include a more realistic modeling of the final 
freeze-out stage \cite{Soff:2000eh}, exploration of fluctuations
in the initial state and ambiguities in the hydrodynamic decoupling 
criterion \cite{Socolowski:2004hw}, viscous effects due to 
incomplete thermalization (i.e. inapplicability of ideal fluid 
dynamics) \cite{Teaney:2003pb}, different (more Landau-type) initial 
conditions leading to strong longitudinal hydrodynamic acceleration
\cite{Renk:2004yv}, and the use of more realistic or different 
equations of state (EoS) for the expanding matter \cite{Huovinen:2005gy}.  
None of these suggestions, individually or in combination, has been 
convincingly shown to be able to solve the HBT puzzle. Motivated by
the blast-wave study in the preceding section, we therefore explore here 
one further possibility: that previous comparisons of the data with 
hydrodynamic models might have been misleading since the RMS variances 
from hydro-generated sources differ significantly from HBT radii 
extracted from a Gaussian parameterization of the correlation function. 
Indications that this is indeed the case have already emerged from the 
work on 1D projections of Hirano and Tsuda \cite{Hirano:2002ds} and 
Kolb \cite{fn1}, and with our new analytic 3D Gaussian fit algorithm 
we can improve on their analysis and study this question in more detail.  

For our study of HBT radii from the hydrodynamic model we use two
different sets of emission functions, obtained from running the 
hydrodynamic code with two different equations of state (EoS). Both
EoS describe the quark-gluon plasma (QGP) as a free gas of massless 
particles, but they differ in their treatment of the late hadronic 
stage when the fireball has cooled below the critical temperature
$T_c{\,\approx\,}165$\,MeV for hadronization. The ``CE EoS'' 
\cite{Sollfrank:1996hd,Kolb:2000sd} assumes that the hadron resonance 
gas remains not only in thermal, but also in {\em chemical equilibrium} 
until final kinetic freeze-out. This fails to reproduce the observed 
hadron yields which correspond to chemical equilibrium at a temperature 
of about 170 MeV \cite{Braun-Munzinger:2001ip}. The ``NCE EoS'' 
\cite{Hirano:2002ds,Kolb:2002ve,Teaney:2002aj} takes the immediate
decoupling of hadron abundances at $T_c$ into account by introducing
{\em non-equilibrium chemical} potentials for each hadron species which 
ensure that the particle yields are held fixed as the temperature and 
density continue to decrease. While the CE EoS was used for the 
hydrodynamic model predictions made for RHIC before the accelerator 
turned on and the hadron abundances were measured, the NCE EoS is more 
realistic and has been used in most hydrodynamic studies since 2002. We 
here explore emission functions obtained with either EoS.

\begin{figure}[t]
\centerline{
\includegraphics[bb=0 20 820 530,width=0.5\textwidth,clip=]%
                {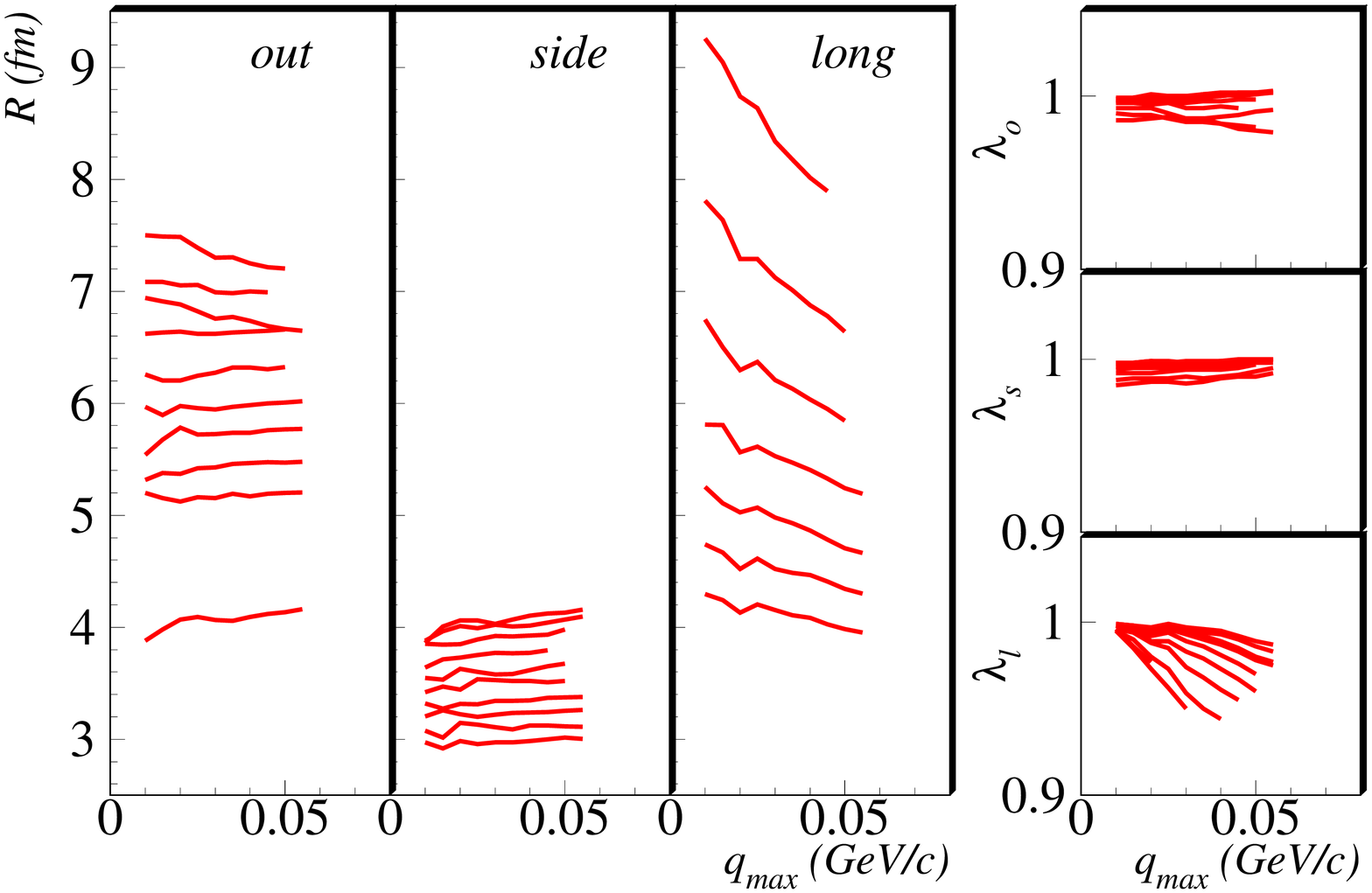}}
\caption{
\label{fig:CEfitrange1Dqmax}
(Color online) 
From the hydrodynamic model with CE EoS, one-dimensional HBT fit parameters 
$R_{\mathrm{1D},i}$ and $\lambda_{\mathrm{1D},i}$ are calculated with Eqs.~(\ref{eq:1Dlammu},\ref{eq:1DRmu}) 
and plotted as a function of the maximum allowed value of any $q$-component; see text   
for details. Each curve corresponds to one of ten values of $K_T$:
0.0, 0.1, 0.2, \dots, 0.9~GeV/$c$. Curves corresponding to high $K_T$ are 
at low (high) values of $R_{\mathrm{1D},i}$
($\lambda_{\mathrm{1D},i}$).
The $R_l$ curves for $K_T{\,\leq\,}0.2$\,GeV/$c$ fall above the plotting range.
}
\end{figure}

\begin{figure}[t]
\centerline{
\includegraphics[bb=0 20 820 530,width=0.5\textwidth]%
                {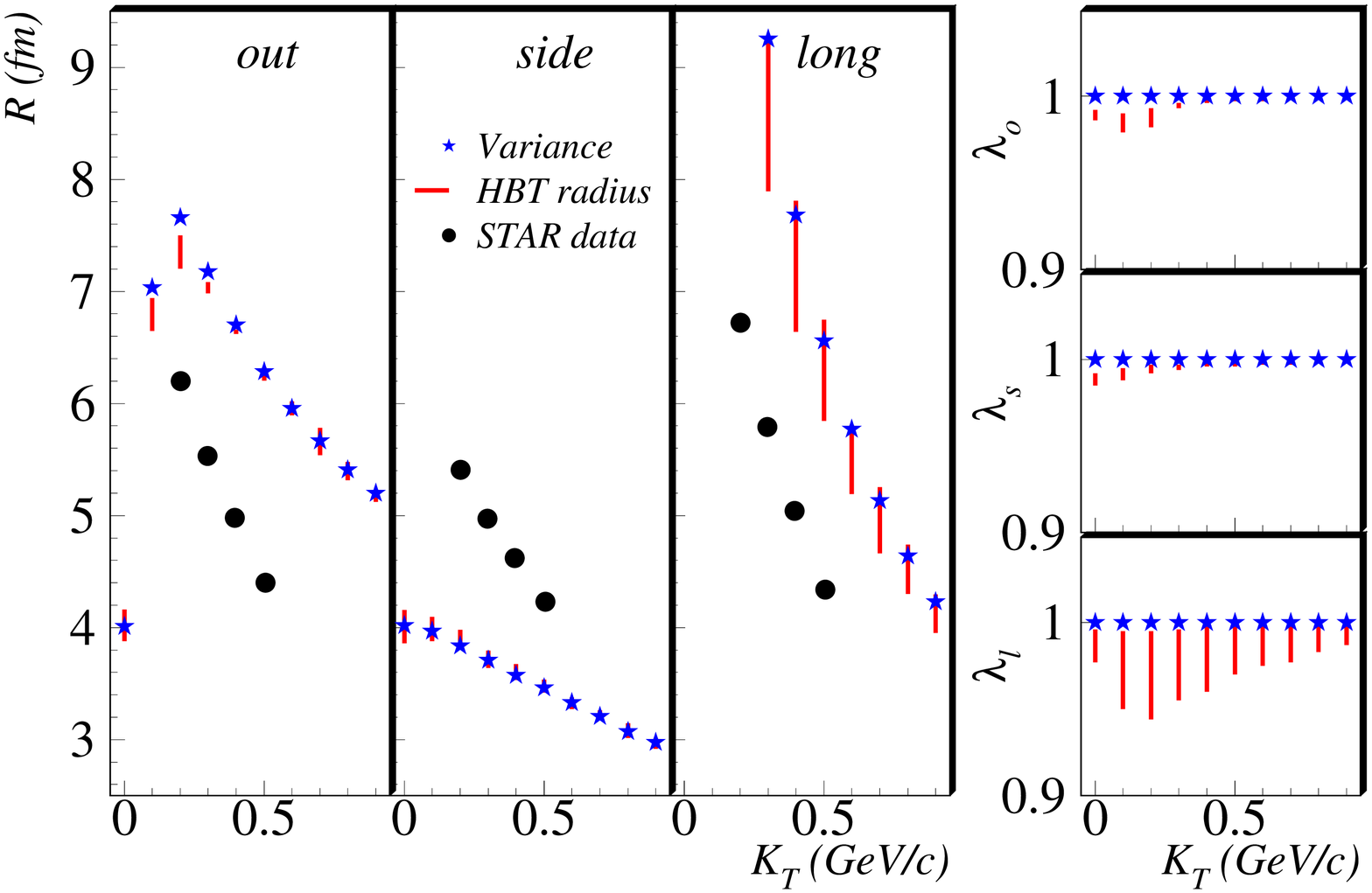}}
\caption{
\label{fig:CEfitrange1DpT}
(Color online) One-dimensional HBT fit parameters $R_{\mathrm{1D},i}$ and 
$\lambda_{\mathrm{1D},i}$ as a function of $K_T$, calculated from the
hydrodynamic model using CE EoS with Eqs.~(\ref{eq:1Dlammu},\ref{eq:1DRmu}).
For a given $K_T$, the vertical red line represents the variation with 
fit range (see Figure~\ref{fig:CEfitrange1Dqmax}). Blue stars represent 
the corresponding radius parameters calculated from the RMS variances 
using Eq.~(\ref{eq:VarianceRadii}). Black circles show STAR data
\cite{Adams:2004yc}, with error bars removed for clarity.
\vspace*{-3mm}
}
\end{figure}

\begin{figure}[t]
\centerline{
\includegraphics[bb=0 20 820 530,width=0.5\textwidth,height=5.58cm,clip=]%
                {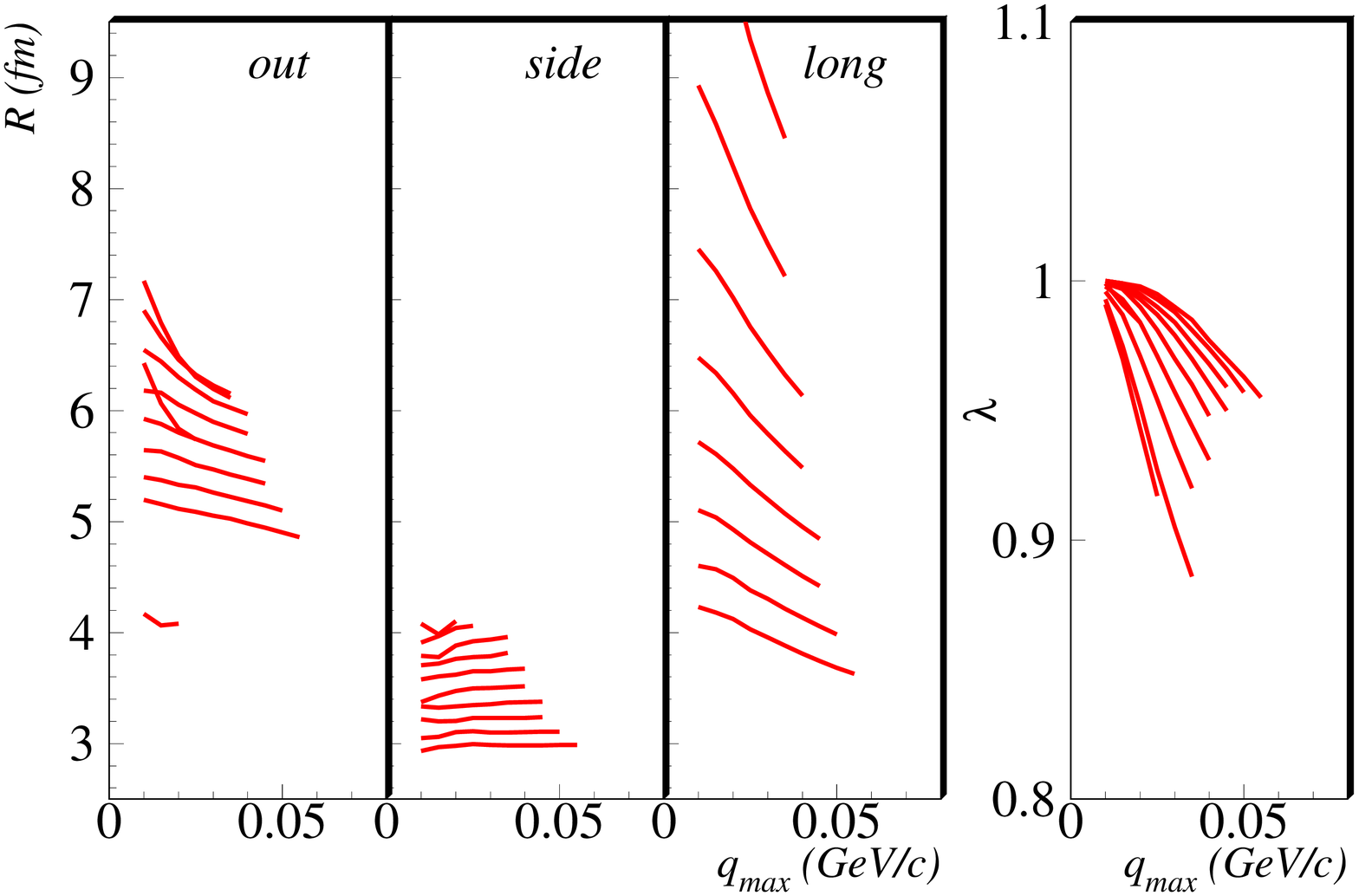}}
\caption{
\label{fig:CEfitrange3Dqmax}
(Color online)
From the hydrodynamic model with CE EoS, three-dimensional HBT fit parameters 
$R_{i}$ and $\lambda$ are calculated with Eqs.~(\ref{eq:LinearEqs}) 
and plotted as a function of the maximum allowed value of any $q$-component; see text   
for details. Each curve corresponds to one of ten values of $K_T$:
0.0, 0.1, 0.2, \dots, 0.9~GeV/$c$. Curves corresponding to high $K_T$ are 
at low (high) values of $R_{i}$
($\lambda$).
The $R_l$ curves for 
$K_T{\,\leq\,}0.1$\,GeV/$c$ fall above the plotting range.
}
\end{figure}

\begin{figure}[t]
\centerline{
\includegraphics[bb=0 20 820 530,width=0.5\textwidth,height=5.58cm,clip=]%
                {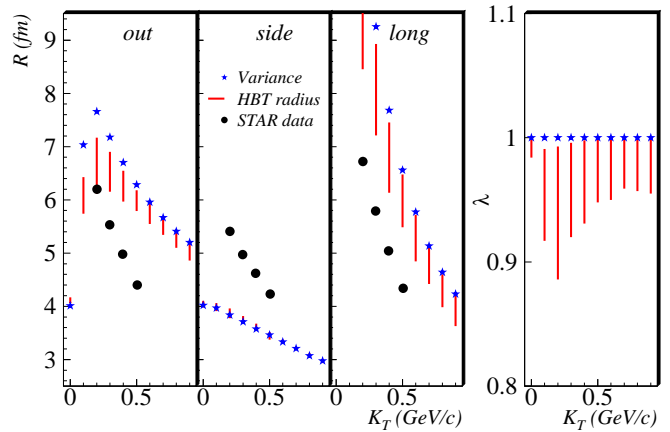}}
\caption{
\label{fig:CEfitrange3DpT}
(Color online) Three-dimensional HBT fit parameters $R_{\mathrm{1D},i}$ and 
$\lambda_{\mathrm{1D},i}$ as a function of $K_T$, calculated from the
hydrodynamic model using  CE EoS with Eqs.~(\ref{eq:LinearEqs}).
For a given $K_T$, the vertical red line represents the variation with 
fit range (see Figure~\ref{fig:CEfitrange3Dqmax}). Blue stars represent 
the corresponding radius parameters calculated from the RMS variances 
using Eq.~(\ref{eq:VarianceRadii}). Black circles show STAR data
\cite{Adams:2004yc}, with error bars removed for clarity.
\vspace*{-3mm}
}
\end{figure}

\begin{figure}[t]
\centerline{\includegraphics[bb=20 20 495 495,width=0.5\textwidth,clip=]%
                            {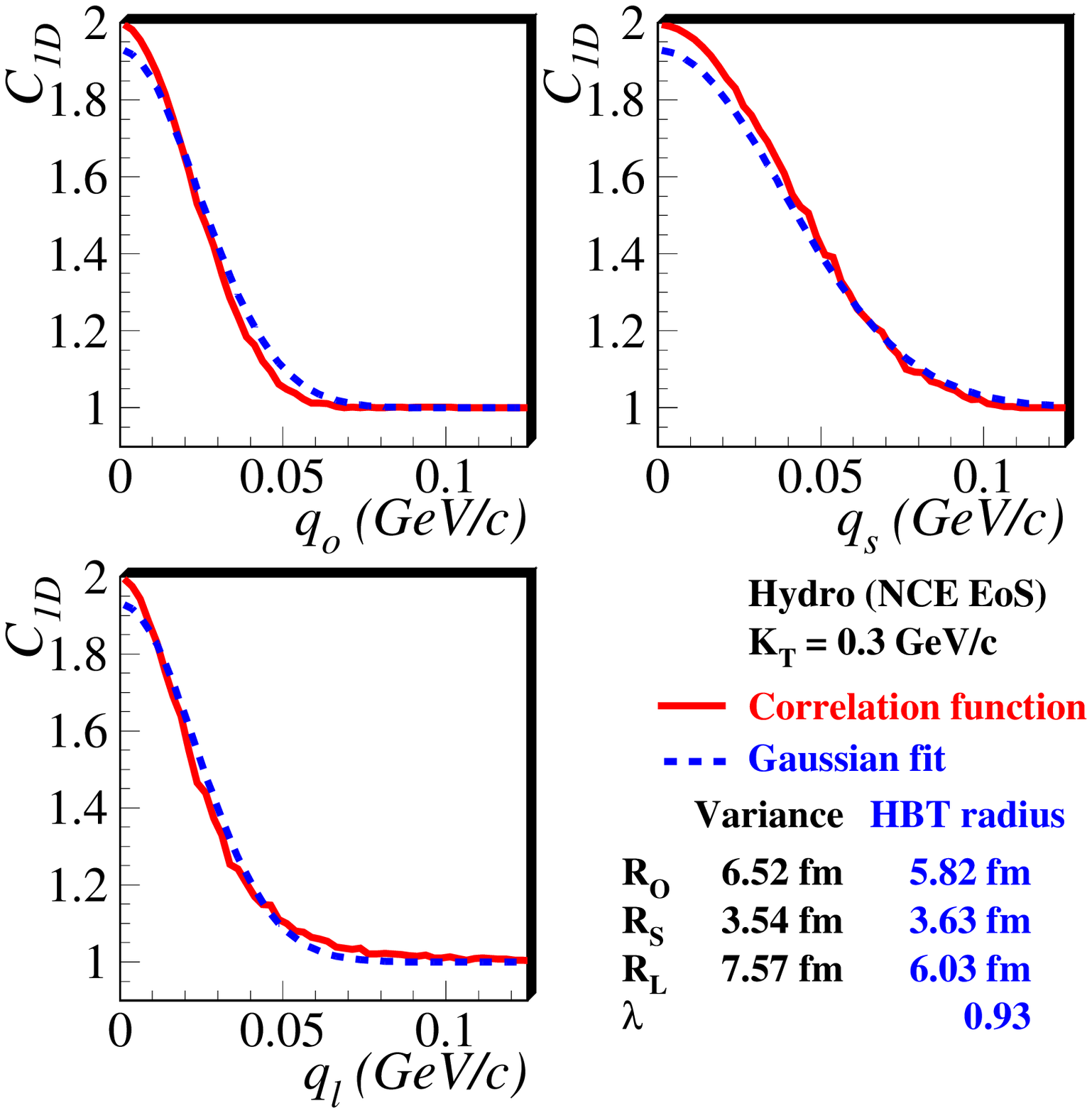}}
\caption{
\label{fig:NCE1dProj-pT0.3}
(Color online) 
Solid (red) curves show
one-dimensional slices of the three-dimensional correlation function calculated
with Eq.~(\ref{eq:TheoryIdOnly}) from the hydrodynamic model with NCE Equation of State, for midrapidity
pions with $K_T=0.3$~GeV/$c$.  Dashed (red) curves show slices of the three-dimensional
Gaussian form of Equation~(\ref{eq:3dGauss}), with ``HBT parameters'' calculated from the analytic 
expressions given in Sec.~\ref{sec3b}.
\vspace*{-3mm}
}
\end{figure}

\begin{figure}[t]
\centerline{\includegraphics[bb=0 20 820 530,width=0.5\textwidth]%
                            {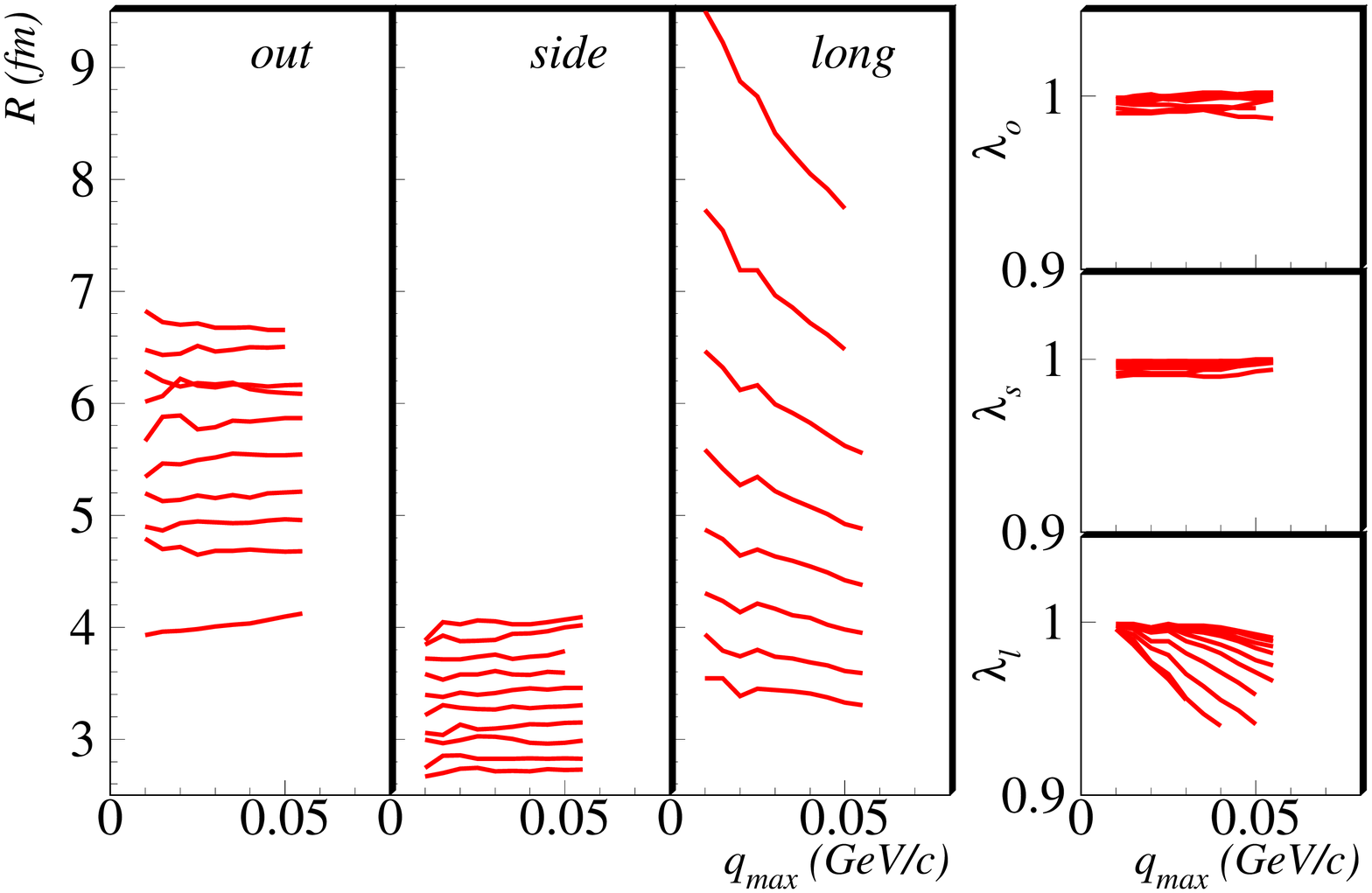}}
\caption{
\label{fig:NCEfitrange1Dqmax}
(Color online) 
From the hydro model with NCE EoS, one-dimensional HBT fit parameters 
$R_{\mathrm{1D},i}$ and $\lambda_{\mathrm{1D},i}$ are calculated with Eqs.~(\ref{eq:1Dlammu},\ref{eq:1DRmu}) 
and plotted as a function of the maximum allowed value of any $q$-component; see text   
for details. Each curve corresponds to one of ten values of $K_T$:
0.0, 0.1, 0.2, \dots, 0.9~GeV/$c$. Curves corresponding to high $K_T$ are 
at low (high) values of $R_{\mathrm{1D},i}$
($\lambda_{\mathrm{1D},i}$).
The $R_l$ curves for $K_T{\,\leq\,}0.1$\,GeV/$c$ fall above the plotting 
range.
}
\end{figure}

\begin{figure}[t]
\centerline{\includegraphics[bb=0 20 820 530,width=0.5\textwidth]%
                            {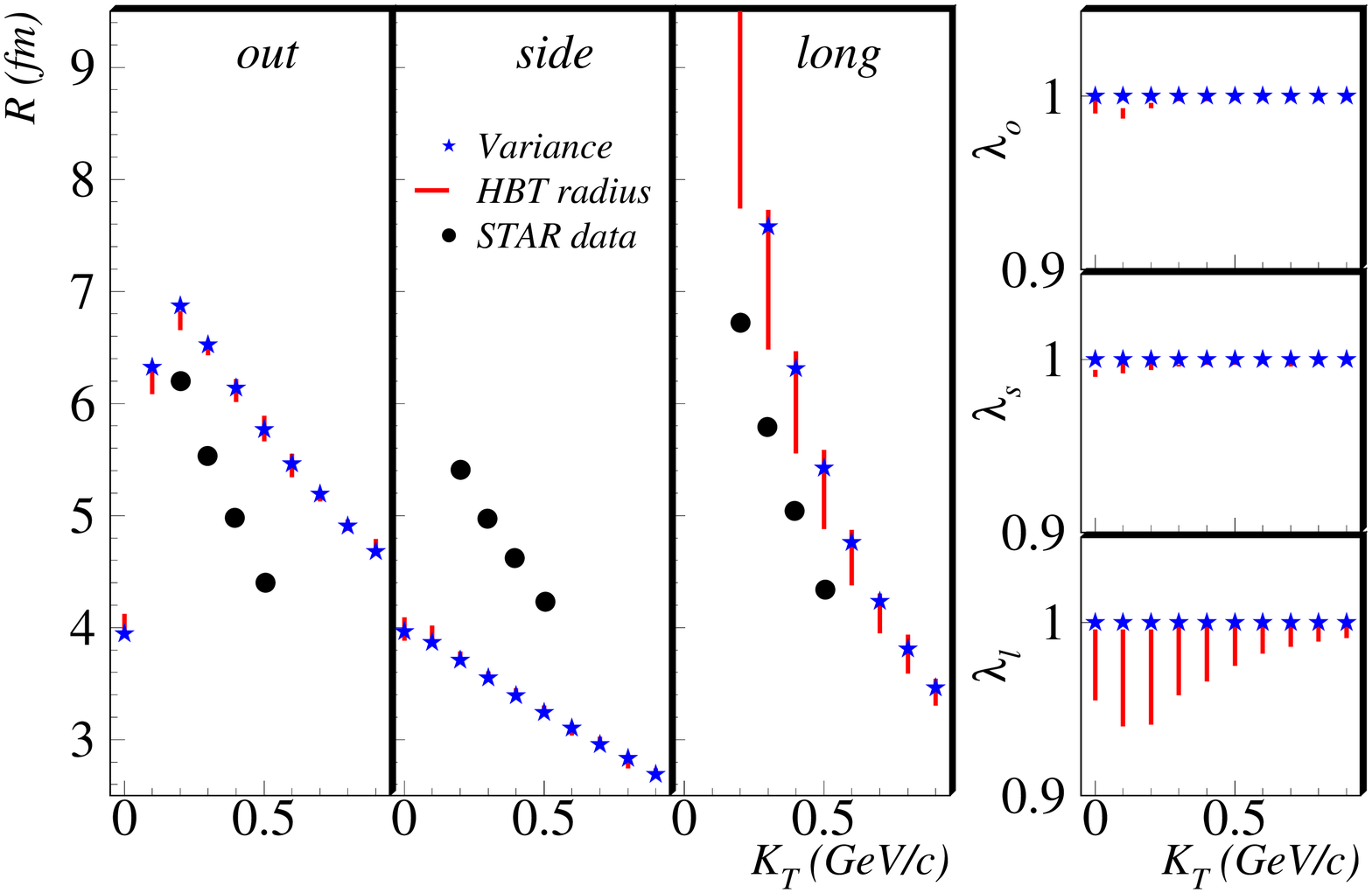}}
\caption{
\label{fig:NCEfitrange1DpT}
(Color online) One-dimensional HBT fit parameters $R_{\mathrm{1D},i}$ and 
$\lambda_{\mathrm{1D},i}$ as a function of $K_T$, calculated from the
hydrodynamic model using NCE EoS with Eqs.~(\ref{eq:1Dlammu},\ref{eq:1DRmu}).
For a given $K_T$, the vertical red line represents the variation with 
fit range (see Figure~\ref{fig:NCEfitrange1Dqmax}). Blue stars represent 
the corresponding radius parameters calculated from the RMS variances 
using Eq.~(\ref{eq:VarianceRadii}). Black circles show STAR data
\cite{Adams:2004yc}, with error bars removed for clarity.
\vspace*{-3mm}
}
\end{figure}

\begin{figure}[t]
\centerline{\includegraphics[bb=0 20 820 530,width=0.5\textwidth]%
                            {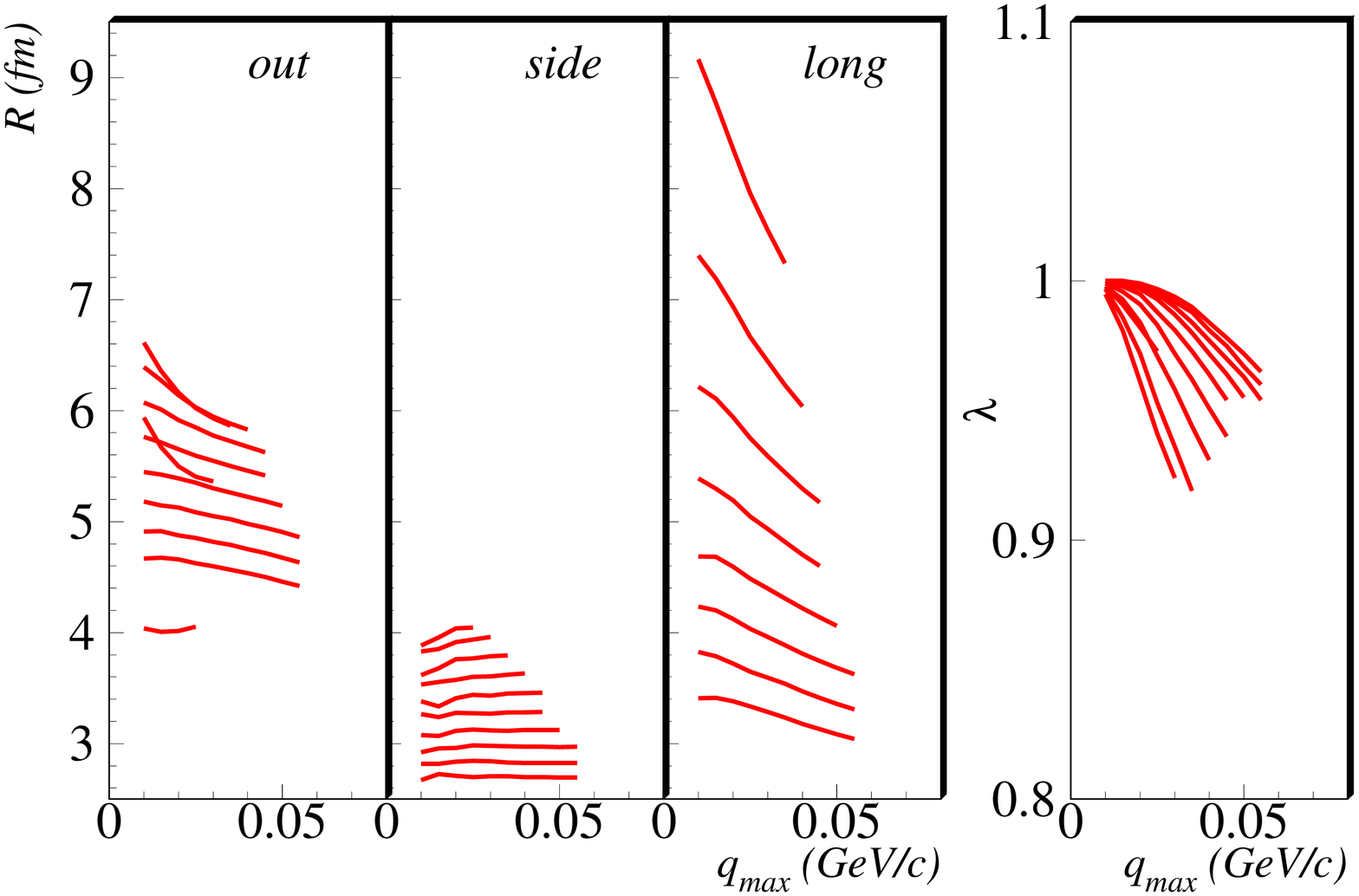}}
\caption{
\label{fig:NCEfitrange3Dqmax}
(Color online) 
From the hydrodynamic model with NCE EoS, three-dimensional HBT fit parameters 
$R_{i}$ and $\lambda$ are calculated with Eqs.~(\ref{eq:LinearEqs}) 
and plotted as a function of the maximum allowed value of any $q$-component; see text   
for details. Each curve corresponds to one of ten values of $K_T$:
0.0, 0.1, 0.2, \dots, 0.9~GeV/$c$. Curves corresponding to high $K_T$ are 
at low (high) values of $R_{i}$
($\lambda$).
The $R_l$ curves for $K_T{\,\leq\,}0.1$\,GeV/$c$ fall above the plotting 
range.
}
\end{figure}

Figures~\ref{fig:CE1dProj-pT0.3}-\ref{fig:CEfitrange3DpT} present 1D 
projections and 1D and 3D fit results, analogous to those from the 
previous section, for the emission function from hydrodynamic calculations 
using the CE EoS. Figures~\ref{fig:NCE1dProj-pT0.3}-\ref{fig:NCEfitrange3DpT} 
show the same for the NCE EoS. Several observations are in order.

As is apparent from Figures \ref{fig:CE1dProj-pT0.3} and 
\ref{fig:NCE1dProj-pT0.3}, the best 3D Gaussian fits do not fully 
reproduce the correlation function, even though the correlation 
function projections themselves appear rather Gaussian. Clearly, 
aspects of the correlation function not apparent in the 
one-dimensional projections are partially driving the 3D fit. 
Further, it is interesting to note that, while the projections in 
the ``side'' direction appear the worst reproduced by the fit, the 
greatest discrepancy between RMS variances and HBT radii are in 
fact in the ``out'' and ``long'' directions (c.f. Figures 
\ref{fig:CEfitrange3DpT} and \ref{fig:NCEfitrange3DpT}). Both of 
these points emphasize that the three-dimensional correlator can
contain important information which does not appear in its one-dimensional 
projections, and thus in the one-dimensional fits. Particularly 
important in this case are strong non-Gaussian features in the 
longitudinal direction which cause a significant suppression of the 
correlation strength parameter $\lambda$ of the 3D Gaussian fit. This
in turn creates the appearance of a ``bad fit'' in the sideward 
direction even though the 1D sideward projection looks quite Gaussian 
itself.

One draws the same conclusion by examining the fit-range systematics. 
As mentioned, non-Gaussian effects generate a variation of the HBT 
parameters with $\qmax$. As seen in Figures \ref{fig:CEfitrange1Dqmax} 
and \ref{fig:NCEfitrange1Dqmax}, fits in the ``out'' and ``side'' 
directions produce 1D radii and directional $\lambda_i$ parameters 
which vary very little with $\qmax$; strong fit-range sensitivity is 
only seen in the ``long'' direction where the 1D projection deviates 
most strongly from a Gaussian form. In the three-dimensional fits, on 
the other hand (c.f. Figures~\ref{fig:CEfitrange3Dqmax} and 
\ref{fig:NCEfitrange3Dqmax}), the strong non-Gaussian features in the 
$q_l$ direction now affect all four fit parameters, generating strong 
fit-range sensitivities also for $R_o$ and $\lambda$.

%
%
%
%
%
%
%

There may (and in general will) be other properties of the 
three-dimensional correlation function to which the 1D projections 
and their Gaussian fits are not sensitive but which affect the 3D 
Gaussian fit. The extracted values for $R_o$ and $R_s$ thus in general 
depend significantly on the detailed conditions under which the 
Gaussian fit is performed. Hence, a meaningful and accurate 
comparison between models and experimental data requires that the
Gaussian fit to the theoretical correlation functions is done under
similar conditions and constraints (e.g. fit range) as the in
experiment.

%
%

\section{Discussion and Conclusions}
\label{sec6}

Let us close with some general observations and summarize our conclusions.

Except inasmuch as it couples HBT radii in a 3D fit, we have not focused 
here on the $\lambda$ parameter, since comparison to measurements of 
$\lambda$ is significantly complicated by experimental artifacts 
\cite{Lisa:2005dd}. This is also the reason why tests of consistency 
between different experiments generally compare HBT radii, not $\lambda$.
In all of the idealized calculations presented in this report 
$C(|\bq|{=}0)\Eq2$, so a purely Gaussian correlation function (generated by 
a purely Gaussian source) would yield $\lambda\Eq1$, with no fit-range 
systematics. Indeed, we find that $\lim_{\qmax\rightarrow 0}\lambda\Eq1$ 
(see e.g. Figure~\ref{fig:CEfitrange3Dqmax}) as expected, but that its 
value declines as more bins are included in the fit. In experimental data, 
several factors cause $\lambda$ to fall below its nominal value of unity. 
Our calculations confirm the generally held folklore that non-Gaussian 
effects may be important to understanding $\lambda$.

Of more fundamental interest are the characteristic length scales of the 
emission region. We have seen that RMS variances of model-calculated 
source functions, which are frequently compared to experimentally 
extracted HBT radii, may systematically differ from ``fitted'' HBT radii 
which characterize the shape of the correlation function from the same
model. Since the latter quantity provides the best ``apples-to-apples'' 
comparison to published experimental data, this can be an important 
observation.

Previous attempts~\cite{Wiedemann:1996ig,fn1,Hirano:2002ds} to estimate the 
effect in hydrodynamical calculations have focused on numerical fits to 
several one-dimensional projections of the calculated correlation function. 
We here presented an analytic method to extract these ``1D HBT radii'' from 
the projections, and further generalized it to the full three-dimensional 
case. The 1D projections represent a set of zero measure of the full 
three-dimensional correlation function and, as we have seen, may not 
be sensitive to important three-dimensional information. This 
information influences the unified three-dimensional fit to the 
correlation function. Since the unified 3D fit most closely mimics 
the procedure of experimentalists, these effects are relevant for
comparisons between models and data.

The magnitude of these effects are model dependent. The non-Gaussian 
nature of emission regions in the blast-wave parameterization has been 
noted before \cite{Retiere:2003kf}. It was shown here to generate only
minor deviations from Gaussian behaviour in the transverse projections
of the correlation function, but the longitudinal projection shows 
significant non-Gaussian features. In a unified 3D Gaussian fit, 
non-Gaussian features were seen to generate fit-range sensitivities 
for all four fit-parameters, leading to significant downward shifts of 
both $R_l$ and $R_o$, especially at low $K_T$, relative to predictions 
based on the spatial RMS variances of the blast-wave source.

\begin{figure}[t]
\centerline{\includegraphics[bb=0 20 820 530,width=0.5\textwidth]%
                            {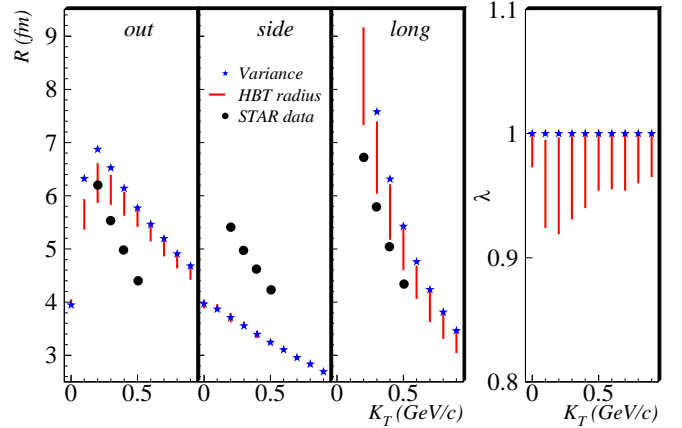}}
\caption{
\label{fig:NCEfitrange3DpT}
(Color online) Three-dimensional HBT fit parameters $R_{\mathrm{1D},i}$ and 
$\lambda_{\mathrm{1D},i}$ as a function of $K_T$, calculated from the
hydrodynamic model using NCE EoS with Eqs.~(\ref{eq:LinearEqs}).
For a given $K_T$, the vertical red line represents the variation with 
fit range (see Figure~\ref{fig:NCEfitrange3Dqmax}). Blue stars represent 
the corresponding radius parameters calculated from the RMS variances 
using Eq.~(\ref{eq:VarianceRadii}). Black circles show STAR data
\cite{Adams:2004yc}, with error bars removed for clarity.
\vspace*{-3mm}
}
\end{figure}

These tendencies were found to be even more strongly exhibited by the
HBT radii extracted from hydrodynamic model sources. The differences 
between HBT radii extracted from 3D Gaussian fits of the correlator
and the values (\ref{eq:VarianceRadii}) calculated from the spatial
RMS variances are quite significant and thus relevant in considerations
of the ``RHIC HBT puzzle''. In particular, for both equations of state 
considered here, the HBT radii in the ``out'' and ``long'' directions
are significantly lower (and closer to the data) than the corresponding 
RMS variances which have been the basis of many ``puzzle'' discussions
(c.f. Figures~\ref{fig:CEfitrange3DpT} and~\ref{fig:NCEfitrange3DpT}).
As in the blast-wave model, these 3D Gaussian fit effects seem to be 
mostly driven by strong non-Gaussian features in the longitudinal 
projection of the correlator.
Combining improvements of using the NCE EoS and the use of HBT 
radii instead of RMS variances brings the hydrodynamic calculations
for the longitudinal radius $R_l$ into fair agreement with the data 
over the entire measured $K_T$ range. A significant improvement is
also seen in the outward direction, but it is mostly concentrated at
low $K_T$, and hence the disagreement between the rather steep 
$K_T$-dependence of the measured $R_o$ radii and the much flatter
$K_T$-dependence of the theoretical results is getting worse.
The fitted sideward radii $R_s$ show practically no deviation from 
the corresponding RMS variances, and the well-known \cite{Heinz:2002un} 
problem that the hydrodynamically predicted values are significantly 
smaller and show much less $K_T$-dependence than the data is not 
alleviated by our improved comparison between theory and data.

%
%
%
%
%
%
%
%

While the results presented here can not offer a resolution of
all aspects of the ``RHIC HBT Puzzle'', they refocus our 
perception of where the most severe problems are located. The 
strong non-Gaussian effects in $q_l$ direction and the resulting 
large downward shift of the fitted longitudinal radii (as compared 
to the corresponding RMS variances) largely eliminate 
the discrepancies between hydrodynamically predicted and measured
$R_l$ values. A number of authors have interpreted the smallness of 
the measured $R_l$ values as evidence for a short fireball lifetime 
$\tau_f{\,<\,}10$\,fm/$c$, inconsistent with the $O(15\,\mathrm{fm}/c)$ 
lifetimes predicted \cite{Kolb:2003dz} by the hydrodynamic model. The
analysis presented here resolves this problem. On the other hand,
even when using the properly extracted Gaussian fit values for $R_s$ and
$R_o$ and after taking into account the resulting decrease of $R_o$
at low $K_T$, the theoretically predicted ratio $R_o/R_s$ is still 
significantly larger than 1 over the entire measured $K_T$ interval,
in contradiction to the data. Furthermore, the decline of both $R_o$ 
and $R_s$ with increasing pair momentum is still much too weak in the 
model, in spite of the large transverse flow generated by the 
hydrodynamic expansion. These aspects of the HBT Puzzle remain 
serious and must be addressed by other theoretical improvements.    

Finally, one should remember that the raw experimental correlation 
functions hardly ever appear very Gaussian, due to additional distortions
by the final state Coulomb interactions between the two charged particles.
Modern methods of extracting the HBT radii from the measured correlator
include these Coulomb effects selfconsistently in the fit function
\cite{Lisa:2005dd}, leading to more complicated (numerical) fit algorithms 
than the analytical one presented in Section~\ref{sec3}. Nonetheless, the 
measured HBT radii extracted from such self-consistent 3D fits are affected 
by non-Gaussian structures in the underlying Bose-Einstein correlations in 
much the same way as discussed here for the simpler case of 
non-interacting particles. Thus, while Coulomb interactions should be 
included in future studies, our analysis should provide a good estimate 
of the direction and magnitude of non-Gaussian effects in blast-wave
and hydrodynamical models, and it points out the importance of such 
effects in the comparison of theory to experiment. 
  
\section*{Acknowledgments}

This work was supported by the U.S. National Science Foundation, grant 
PHY-0355007, and by the U.S. Department of Energy, grant 
DE-FG02-01ER41190.


\begin{thebibliography}{99}
\bibitem{fn2}
  Note that for one-dimensional correlations measured in 
  $Q_\mathrm{inv}\Eq\sqrt{-q^\mu q_\mu}$, 
  as opposed to one-dimensional slices of the three-dimensional correlator, 
  phase-space considerations usually produce larger uncertainties at 
  small $Q_\mathrm{inv}$.
\bibitem{fn3}
  When the HBT radii are large (e.g. at low $K_T$), the 
  correlator approaches unity quickly with increasing $|\bq|$, and the 
  quantity $\ln\bigl[\Cqk{-}1\bigr]$ becomes numerically unwieldy; in 
  these cases only small values of $\qmax$ are used.
\end{thebibliography}
\end{document}